\documentclass[aps,prb,superscriptaddress,reprint,
notitlepage]{revtex4-2}
\usepackage{amsmath}
\usepackage{graphicx} 
\usepackage{physics} 
\usepackage[dvipsnames]{xcolor}

\DeclareMathOperator\erfc{erfc}
\newcommand{\mctwo}{Department of Microtechnology and Nanoscience -- MC2,
Chalmers University of Technology, SE-41296 Gothenburg, Sweden}
\newcommand{\ecom}{e-Commons, Chalmers University of Technology, SE-41296 Gothenburg, Sweden}
\newcommand{\donostia}{Donostia International Physics Center (DIPC), 20018 Donostia/San Sebastian, Spain}
\newcommand{\chalchem}{Department of Chemistry and Chemical Engineering,
Chalmers University of Technology, SE-41296 Gothenburg, Sweden}

\begin{document}

\title{Stress-based structure optimization for a range-separated hybrid\\ 
van der Waals density functional}

\author{Per Hyldgaard}
\affiliation{\mctwo}
\email{hyldgaar@chalmers.se}
\author{Yunqi Shao}
\affiliation{\ecom}
\author{Raul Quintero-Monsebaiz}
\affiliation{\mctwo}
\affiliation{\donostia}
\author{Lars {\"O}hrstr{\"o}m}
\affiliation{\chalchem}

\date{\today}

\begin{abstract}
Complex matter, often partly soft and pliable, is the generic form of material systems. We must therefore generally call on density functional theory (DFT) to first predict the atomic structure before we can use it to also characterize (expected) properties. A recent range-separated hybrid 
(RSH) van Waals density functional (vdW-DF), `vdW-DF2-ahbr' 
(abbreviated AHBR) [PRX \textbf{12}, 041003 (2022)],
shows promise as a high-accuracy predictor of both binding energies and
structure, from molecules to solids.  However, the present implementation in Quantum Espresso (QE) does not, in our experience, support robust stress-based unit-cell optimization of RSHs. Here we document and illustrate work enabling practical 
AHBR-based complex-matter discovery: We port the AHBR XC functional to the Vienna Ab Initio Simulation Package (VASP) where use of stress-based DFT optimization is already stable also for RSHs.  We test the implementation by comparing 
high-accuracy AHBR-QE and AHBR-VASP predictions of non-covalent 
molecular interactions and for structure and cohesion of simple 
bulk structures. We also illustrate and test the use of the new 
AHBR-VASP implementation to predict and understand
(atomic and anti-ferromagnetic) structure in distorted-rocksalt metal monoxides. Finally, we illustrate 
use for complex-soft-matter discovery 
by predicting motifs for layer stacking  
in the C2N  covalent-organic framework (COF) system [Nat.\ Commun.\ \textbf{6}, 6486 (2015)].   
\end{abstract}

\maketitle

\section{Introduction}

Modern materials and chemistry research as well as technology development challenge first-principle density functional theory (DFT) \cite{hoko64,kosh65,GKSstart,BurkePerspective,beckeperspective,burke}.
Improved  function is often set by details in the ground-state atomic and electronic structure, as reflected in reactions and catalysis \cite{JKNBILloss1979,BILsurfReact1979,AdsAspectBIL1991,HeadGordonTully95,Noblest,KroesScatter2008,Diaz2009,catalysisvdW15,KroesDC2021,SBH17,DiscoverCatalysts2020,DiscoverBimetalCatalysts2021,Cu2O2,AuerbachScatter24,H2COinteract2026,Leiden26} or expressed in the coupling to vibrational excitations
\cite{Baroni2001,Sabatini2013p041108,ErhHylLin15,BrownAltvPRB16,PeGrRo20,jewahy20}. There is also interest in completing mutually consistent atom- and electronic-structure predictions  to develop better optoelectronics \cite{Hedin65,FW7,la70,helujpc1971,Hedin80,AuJoWi00,cococcioni2005,dabo2010,kronik2012,KraKro13,OTRSHalga,ferretti2014,nguyen2015,nguyen2016,RanPRB16,OTRSHadsorb17,colonna2018,nguyen2018,Elliott2019,jewahy20,WiOhHa21,gennaro2022,AHBRlaunch,colonna2022,ChiDFT23,GoGaOh2024,ChiDFT24,AHBRmRSH25,NitrogenBasesAHBR-mRSH26,H2COinteract2026,hBN2026,DiscoverFerroelectrics2026}. 

For example, theory-led or theory-assisted high-throughput 
screening of candidate systems \cite{GBRV,DiscoverBimetalCatalysts2021,AutonomousOpt2024} is a strong driving force for modern 
materials discovery, i.e., a search for optimizing or just realizing a given desired property or 
function, for example, Refs.\ \cite{Yasher1999,2001surfscience,Sofo07,YaYuSuNe17,DiscoverCatalysts2020,DiscoverBimetalCatalysts2021,AutonomousOpt2024,2DdiscoverySC25,DiscoverFerroelectrics2026}. Such exploration is no longer limited to picking among what are existing, if at times only partially understood materials, composites, and systems. Discovery should  include new ideas along with assessments of whether such virtual materials design is worthy of actual synthesis or suitable functionalization \cite{Sofo07,RahmBrinck2010,Frostenson2024,2DdiscoverySC25}. A common characteristics of modern materials discovery is therefore that we must 
understand complex matter, i.e., systems for which we do not already have trusted experimental structure characterization; 
This is, in fact, the typical state of matter and systems
\cite{langrethjpcm2009,Berland_2015:van_waals,AHBRlaunch,Hard2Soft,Cu2O2,EUrecommendation}, yet we must still predict and compare the function among all candidates for progress \cite{Yasher1999,Carlo2003,RaRuHyLu03,Rohrer2010a,Rohrer2010b,RahmBrinck2010,rohrer11p165423,YaYuSuNe17,shukla2018borophene,shukla2019modelling,umrao2019anticarcinogenic,shukla2020rectifying,PeGrRo20,MOFdobpdc,Hard2Soft,shukla2022nt,Frostenson2024,SeSoGoBe24,HybridPerov26}. 

With such theory-guided materials discovery it is also motivated to develop highly-accurate exchange-correlation (XC) functionals for general-purpose \textit{ab initio} DFT and to make it easier to use such new DFT 
for efficient structure optimizations in general types of (heterogeneous) materials problems. The actual synthesis stage remains the costly step 
but one need not go there until a general high-throughput theory 
screening, e.g., Refs.\ \cite{YaYuSuNe17,WaFeLi25,PoRiNe26,2DdiscoverySC25,DiscoverFerroelectrics2026}, 
is itself validated within unbiased theory. Accordingly, we can accept a computationally more expensive last-DFT step, for example, being a hybrid with some Fock-exchange component and thus residing within the so-called generalized Kohn-Sham (KS) DFT framework \cite{GKSstart}. However, the chosen (hybrid) XC functional for theory validation should be recognized for its general-purpose applicability and systematic accuracy. Also, the underlying discovery purpose suggests picking here an 
XC choice without adjustable parameters: When DFT breaks new ground, there  may not be enough trusted data to guide machine-learned functionals to what should be broadly-trusted accuracy \cite{EUrecommendation}.  

One strategy for trusted XC designs involves building exclusively from generic physics input, including 
many body perturbation theory (MBPT), conservation laws, screening, and scaling insight \cite{bohrlindhard,mabr,gulu76,lape77,ma,ra,lavo87,pewa92,anlalu96,dobdint96,pebuer96,pebuwa96,Dion,thonhauser,JPCMreview}. Importantly, we can use these XC ideas also in corresponding hybrid forms \cite{GKSstart,BeckeIII,Gorling93,Perdew96,Burke97,PBE0,HSE03,HJS08,JiScHy18b,DefineAHCX,AHBRlaunch}. We note that the formal logic of DFT implies that accuracy for the electronic structure must directly reflect the atomic structure \cite{hoko64,kosh65,GKSstart,FW7}. We also note that by staying within a MBPT-guided XC framework, we make
it plausible that we maintain mutual consistency \cite{FW7,gulu76,lape77,GKSstart,Dion,thonhauser,nguyen2015,RanPRB16,OTRSHadsorb17,JPCMreview,AHBRlaunch,colonna2022,AHBRlaunch,ChiDFT23} and enable independent validation in terms of, e.g., photoemission 
energies and signatures \cite{OTRSHalga,OTRSHgap,ferretti2014,nguyen2016,OTRSHadsorb17,WiOhHa21,NitrogenBasesAHBR-mRSH26,hBN2026}. For a specific problem, semi-empirical and empirical (fitted) XC functional may perform better at similar computation costs, especially for molecules
\cite{gmtkn55}. However, in general material discovery we typically face both solids, surfaces, and heterogeneous (solid-molecule) interfaces. We like the physics-guided strategy for continued XC development that can be trusted for both sides of a molecule-solid interface, for example, as discussed in Refs.\ \cite{DefineAHCX,AHBRlaunch}. This is because this strategy
may help minimize the bias towards some types of 
composites of materials. In fact, one of us has long worked to further develop the van der Waals density functional (vdW-DF) method \cite{lavo87,anlalu96,ryluladi00,rydberg03p126402,Dion,thonhauser,lee10p081101,behy14,hybesc14,Thonhauser_2015:spin_signature,Berland_2015:van_waals,JPCMreview,DefineAHCX,AHBRlaunch,AHBRmRSH25} to become better suited for an unbiased, yet accurate
material exploration \cite{bearcoleluscthhy14}.

\begin{figure}
    \centering
\includegraphics[width=0.34\linewidth]
{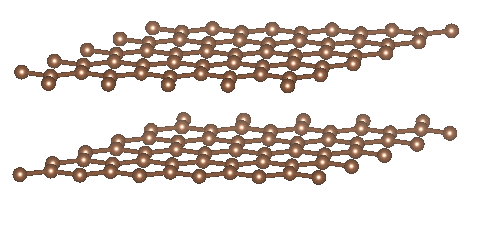}
\includegraphics[width=0.34\columnwidth]
{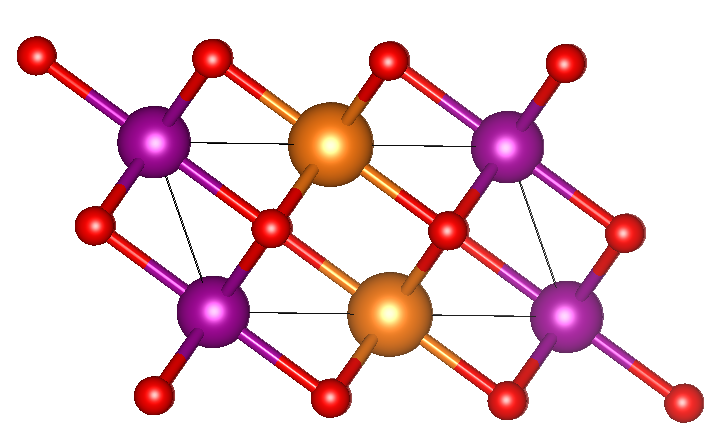}
\includegraphics[width=0.21\columnwidth]
{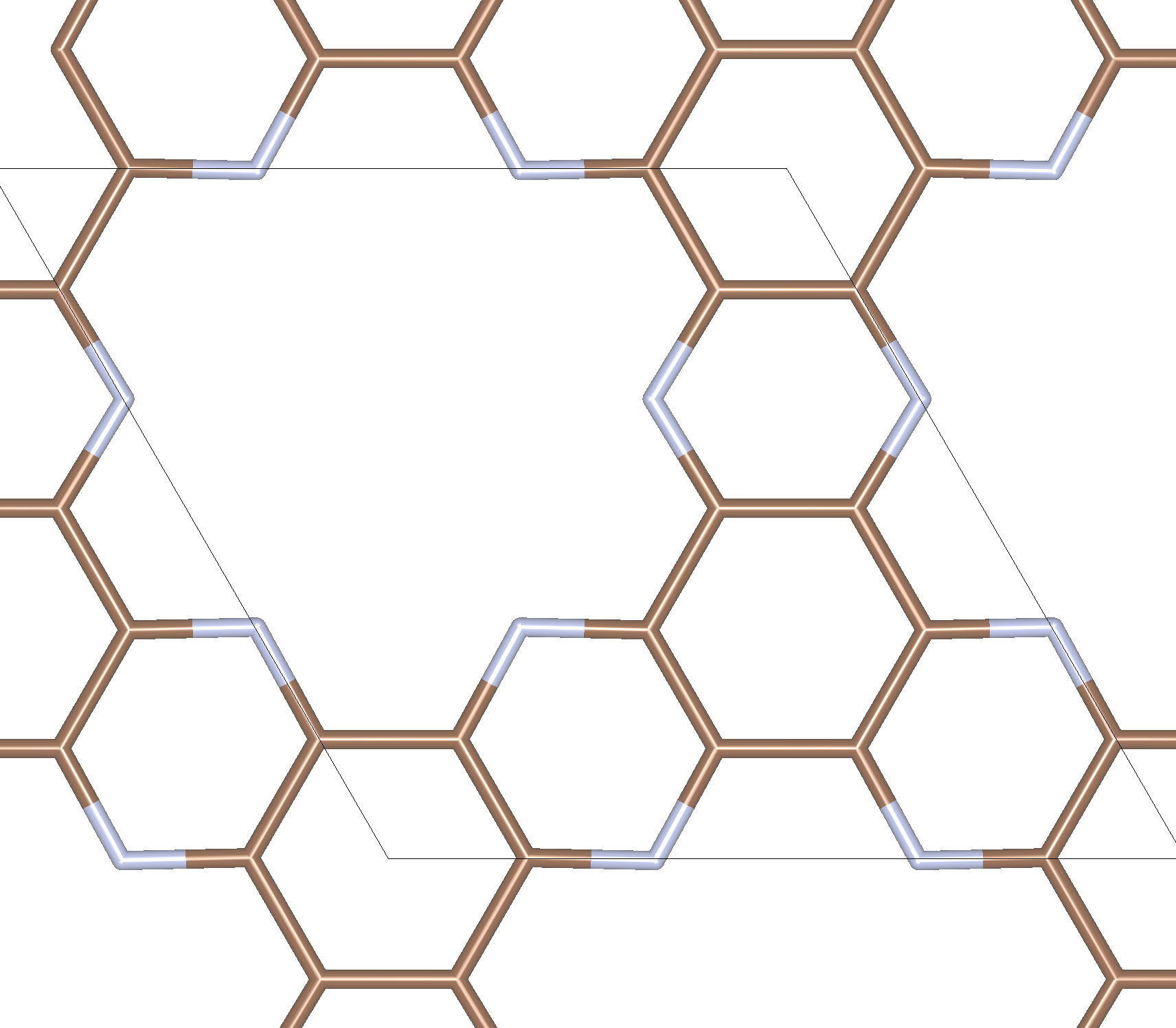}
\includegraphics[width=0.90\linewidth]{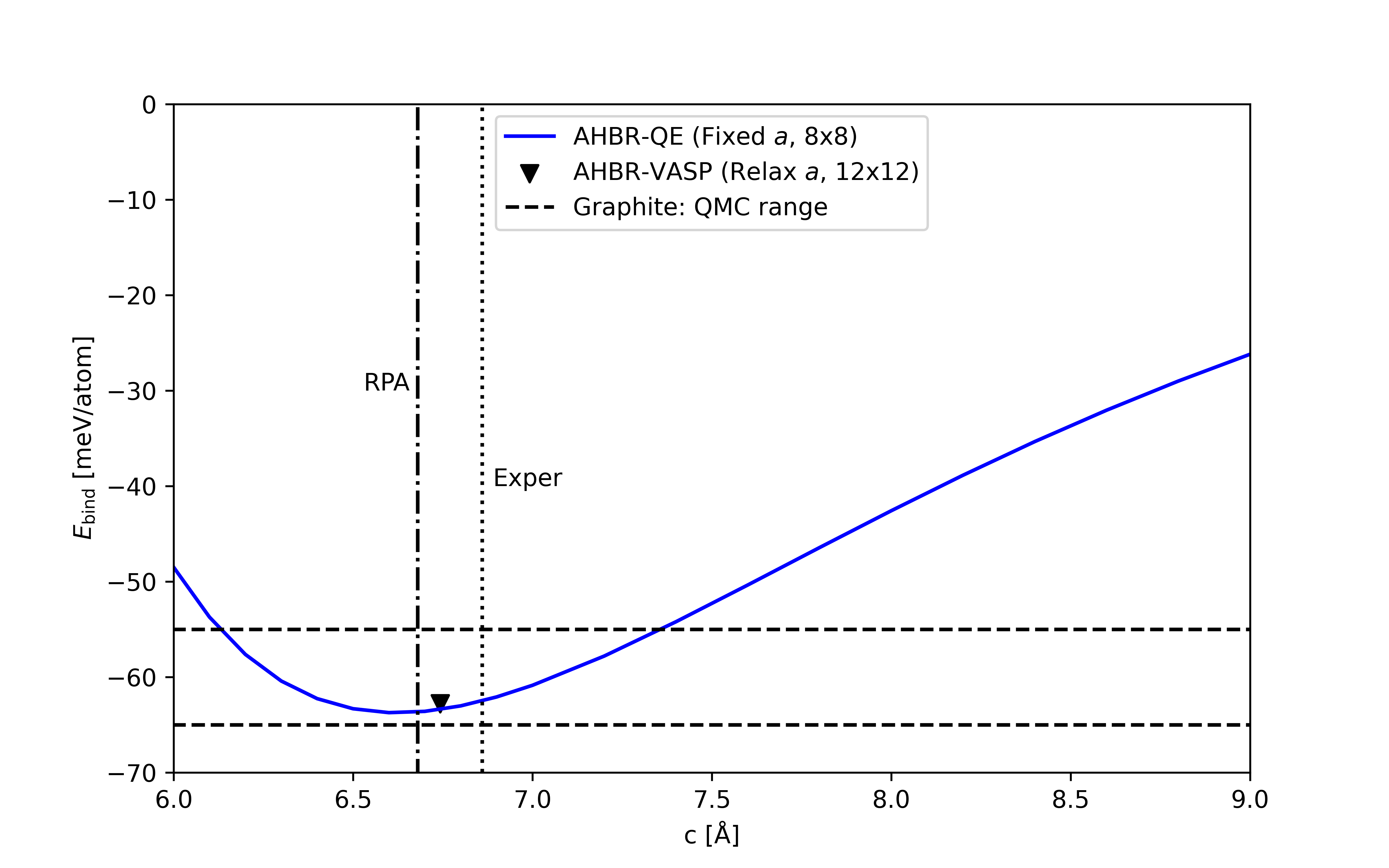}
    \caption{\textit{Top:} Schematics of the atomic structure in three examples  
    of here-discussed materials, layered hexagonal for 
    bulk graphite (left),  slightly distorted rocksalt for bulk MnO (middle), and 
    incompletely-characterized stacking of layered sheets for `C2N' COF \cite{C2Nstart2015} (right). For the first two examples, we use brown and red balls to represent 
    carbon and oxygen atoms, respectively, and magenta (gold) balls to represent 
    Mn atoms with a predominant spin-up (-down) configuration. For C2N-COF, where there is no conclusive experimental nor theory data on the layer stacking, we use instead brown and aquamarine vertices to represent C and N atoms as organized within the individual, strictly flat, C2N sheets; We are here using AHBR-VASP to discover motifs of the layer assembly into the actual (bulk) COF. \textit{Bottom:} Contrast 
    of the AHBR-QE and AHBR-VASP characterizations for graphite. For a computationally tractable structure determination by AHBR-QE, we fix the in-plane lattice constant $a$ across all assumed $c$ values, compute the system energy $E_{\rm bind}$ (curve) relative to the value computed at $c\to \infty$, and manually identify 
    the minimum. In the AHBR-VASP determination a single run with self-adjusting unit-cell optimization suffices to set the structure (triangle), with the energy prediction again requiring a subtraction of the large-$c$ reference energy. Comparison with the pair of horizontal dashed lines representing spread in literature QMC result for $E_{\rm bind}$  \cite{Spanu09p196401,GaKiPa2014,MoDrFa2015,ShKiLe2017}  (and with the vertical dotted and dashed-dotted lines for experimental and RPA characterizations for $c$ \cite{Lebegue10,OlTh2013}), shows that both of these AHBR-based schemes work for this simple material, in spite of the approximation made for the AHBR-QE approach. However, the ABHR-QE approach becomes intractable already for the slightly-disturbed rocksalt cases.
    }
\label{fig:SchematicsVCrelax}
\end{figure}

Here we seek to convert a promising XC-candidate 
for accurate, general-purpose (parameter-free,
MBPT-guided) atom- and electronic-structure 
predictions into a more practical tool for helping 
discovery. Our work takes off from the
observation that a recently defined
range-separated hybrid (RSH) vdW-DF, namely
the vdW-DF2-ahbr \cite{AHBRlaunch}
is found robust in making accurate materials predictions from molecular to traditional bulk systems.  The naming, vdW-DF2-ahbr (abreviated AHBR), highlights that it is crafted off of both the vdW-DF2 \cite{lee10p081101}
and an analytical(-exchange)-hole modeling \cite{HJS08,DefineAHCX,AHBRlaunch} of Hamada's revised-Becke86 \cite{becke1986p7184,hamada14}.  
The AHBR is presently exclusively implemented and released in \textsc{QuantumESPRESSO} 
\cite{QE,Giannozzi17} (QE)
which is itself a widely 
trusted planewave-DFT code. There we  use AHBR with a class of electron-rich optimized norm-conserving Vanderbilt \cite{ONCV,sg15} (ONCV)  pseudopotentials
(PPs). This AHBR-PP-QE approach is both robust and  accurate on general molecular problems, for example, as compared to both dispersion-corrected HSE and B3LYP \cite{BeckeIII,LYP} on the set of 5 subsets of the GMTKLN55 benchmark suite and on simple solids \cite{AHBRlaunch}.

However, our present AHBR-QE coding is not set up 
to enable generic material-discovery use. This
is because hybrid-XC studies in  QE presently have limited stability when used with projector-augmented wave (PAW) potentials (that can potentially lower the computation costs and increase accuracy) or with stress-based variable cell optimization 
\cite{Nielsen1985,sabatini12p424209,Hard2Soft}. For complex matter cases, we are presently forced to first create a database of fixed-unit-cell runs and then complete the structure search on that database, much like the situation faced at the start of the wider vdW-DF program \cite{ziambaras03p064112,ziambaras07p155425}. With access to stress-based optimization, we can significantly accelerate structural predictions by AHBR, especially when facing complex matter with many unit-cell parameters \cite{RanPRB16,Hard2Soft}.

In practice, our here-reported enhancement of AHBR to an effective tool for unbiased discovery takes the form of porting this RSH vdW-DF from QE to 
an in-house implementation in the Vienna Ab initio Simulation Package,
(VASP). We thus create also an AHBR-PAW-VASP tool for materials
predictions and modeling. This enhancement is detailed here 
while we also demonstrate that 1) we retain, and in some cases improve, the AHBR accuracy 
with the change from ONCV PPs to PAW setups and that 
2) the VASP implementation permits us to
complete more complex structure searches. 

The paper is organized as follows. The following two sections 
present the theory and code implementation as well as computational details. Section IV contains our results and 
discussions including documentation for implementation 
robustness and demonstrators of use for complex-matter
characterization and for discovery. The paper has two 
appendices, comparing details in AHBR-VASP and AHBR-QE descriptions of noncovalent interactions in molecules and documenting convergence in AHBR-VASP and AHBR-QE studies of bulk systems. 

\begin{table*}
\caption{\label{tab:INCARset} VASP `INCAR' settings unique to using 
AHBR and HSE06 for variable-cell structure optimization in VASP.}
\begin{tabular}{l|cc|  l}
\hline
\hline
INCAR PARAM & HSE06 & AHBR & Comment (When relevant: Split for HSE/AHBR cases) \\
\hline
\verb|LHFCALC| & T & T & Compute Fock-exchange component, for use in, e.g., HSE/AHBR RSH XC functionals.\\
\verb|HFSCREEN| & 0.2 & 0.2 & Inverse length scale $\gamma$ ({\AA}$^{-1}$) for extent of Fock-exchange mixing.\\ 
\verb|GGA| & OMIT & MK & Set by PAW input/As for B86R (short for rev-vdW-DF2).\\
\verb|PARAM1| & OMIT & 10/81 & -/As for B86R (rev-vdW-DF2).\\
\verb|PARAM2| & OMIT & 0.7114 & -/As for B86R (rev-vdW-DF2).\\
\verb|AGGAC| & OMIT & 0.0 & Uses default/As for all vdW-DFs. \\
\verb|LUSE_VDW| & OMIT (F) & T & -/As for all vdW-DFs\\
\verb|ZAB_VDW| & OMIT & -1.8867 & -/Set as in vdW-DF2 and B86R (rev-vdW-DF2).\\
 \hline
\verb|LASPH|  & T & T & Non-spherical contributions, by RSH and by vdW-DF needs.\\
\verb|NELMDL|  & -4/-8 & -4/-8 & Suggested for speed. Same logic is used in QE.\\
\verb|ISIF|  & 3 & 3 & Also 4 (7) when enforcing bilayer geometry (true-RS symmetry).\\
\hline
\hline
\end{tabular}
\end{table*}

\section{Theory and code implementation}

The prospect of using an accurate, general-purpose
XC functional for (reasonably-fast) extended-materials predictions is today heightened by computationally efficient implementations of XC for generalized-DFT (having a truly nonlocal exchange) and vdW-DFs (having a truly nonlocal correlation component) \cite{HSE03,Dion,dionerratum,thonhauser,HJS08,roso09,HSEsol,linACE,DefineAHCX,AHBRlaunch}. The combination, i.e., full non-locality on both parts of the XC description, is implemented in QE \cite{PaoloElStruct1,DefineAHCX} and (with this work) in VASP. All such XC-energy descriptors remain a functional of the electron density, denoted $n(\mathbf{r})$. The popular PBE version \cite{pebuer96} 
of the generalized gradient approximation \cite{rasolt,lape80,lameprl1981,lavo87,lavo90,pewa86,pewa91,pebuer96,PBEsol} (GGA) is often a robust XC functional for use in traditional DFT, i.e., when staying in the KS framework and using just a semilocal form in the XC specification \cite{pewa86,pebuer96}. It can be a good choice for material discovery but 
only when we are sure that the search will stay
within the class of matter with dense $n(\mathbf{r})$ variations, i.e., hard (bulk) materials with strong bonding in all three dimensions. However, for unbiased discovery, we should not \textit{a priori} limit ourselves regarding the $n(\mathbf{r})$ variation.

A set of recent regular (that is, non-hybrid) vdW-DFs eclipse the accuracy and versatility of the PBE, generally and often also in the case of the hard-matter problems \cite{Gharaee2017,JPCMreview,DefineAHCX,AHBRlaunch}. This is true even if these vdW-DFs have just a GGA-type exchange. PBE can fail
significantly for heterogeneous systems, and 
in particular for systems that have 
important regions with a sparse electron density $n(\mathbf{r})$. Here `important region' is taken to mean one of key consequence for binding or function \cite{langrethjpcm2009,Berland_2015:van_waals}.
For example, PBE certainly fails when it comes to inter-molecular interactions problems and is less accurate on (transition) metals and simple surface properties \cite{Gharaee2017,JPCMreview,AHBRlaunch} than, e.g., the consistent-exchange vdW-DF-cx \cite{Dion,behy14,bearcoleluscthhy14} (abbreviated CX) and rev-vdW-DF2 \cite{lee10p081101,hamada14}, here denoted vdW-DF2-ahbr (B86R) to emphasize the
nature of its GGA-type exchange \cite{becke1986p7184,hamada14,AHBRlaunch}. 

Further accuracy gains generally arise, across many types of systems, when we move to hybrid forms of versatile vdW-DFs, including the AHBR that is build as a RSH vdW-DF \cite{DefineAHCX,AHBRlaunch,AHBRmRSH25} off of
B86R. The AHBR design work largely follows the crafting of HSE06 \cite{HSE03}, namely the standard RSH GGA based on the popular PBE \cite{pebuer96,EP98,HSE03,HJS08}. As general-purpose predictors, we find that more trust can be ascribed to AHBR than to PBE or HSE; The same holds when comparing AHBR to the dispersion-corrected \cite{grimme3}
HSE+D3, see 
Ref.\ \cite{AHBRlaunch}. 

Excitingly, the use of both of these RSH forms, HSE and AHBR, are today practical also 
for extended matter and even for studying adsorption at transition-metal surfaces. 
This follows because there now exists efficient implementations \cite{KresseNJP,jarlkresse,PaoloElStruct1} of the so-called adaptively compressive exchange (ACE) formulation of Fock exchange \cite{linACE}, in both \textsc{VASP} and in QE. 
One can, for example, use AHBR-PP-QE  \cite{AHBRlaunch} to complete RSH vdW-DF characterizations of CO adsorption on noble-metal and Pt surfaces \cite{DefineAHCX,AHBRlaunch,Leiden26,H2COinteract2026}. Convergence and accuracy at such problems require sampling across very many in-surface $k$ points  and differences $q$ in the Fock-exchange evaluations underpinning our hybrid-DFT studies.

At the same time, it is only the VASP implementations of the accelerated 
(extended-system) Fock-exchange 
evaluation that also permits robust stress-based structure optimization with such RSH functionals.
This status motivates us to now port the AHBR  to VASP.

\subsection{Relevance of stress-based AHBR optimization}

The top row of panels in  Fig.\ \ref{fig:SchematicsVCrelax} show schematics 
of graphite, of what proves to be \textit{almost}-rocksalt 
antiferromagnetic-type-II MnO, and of individual-sheet atomic configuration of the C2N COF system \cite{C2Nstart2015}. We note that the three-dimensional structures of the
latter two insulating/semiconducting materials are incompletely characterized in experiments. We therefore need to start by a parameter-free theory characterization of structure when we also seek to predict and understand their properties. The same is certainly true when we furthermore seek exploration, i.e., when we consider them as a potential class for broader discovery. The latter two cases are simple examples that  motivate the here-presented implementation of AHBR-VASP. More broadly, stress-based structure optimization makes discovery practical.

We first observe that we can use and are using our present AHBR-QE coding to predict the behavior for what we denote as structurally simple problems. By `structurally simple' we mean cluster-type problems and extended-materials cases where atomic relaxations can be expected to occur essentially as decoupled from the task of also predicting the correct unit-cell organization. We can also address problems where it suffices to complete intra-cell atomic optimization in a sweep of fix-cell studies over no more than a single cell parameter. For example, we use AHBR-QE to predict structure and hence properties in molecules and for most molecule-molecule interaction problems \cite{AHBRlaunch,NitrogenBasesAHBR-mRSH26}, but we cannot presently address general molecular crystals directly in AHBR, since interactions can there
drive deformations of the unit cell form \cite{RanPRB16}. 
Also, we can use AHBR-QE to predict the relaxations, energies, and dynamics of surface or internal-surface adsorption problems \cite{AHBRlaunch,Cu2O2}
as long as the unit-cell nature of the substrate 
has already been understood; This is possible 
(at least at low coverages) because the molecule 
adsorption is not generally causing more than local relaxations \cite{MOFdobpdc,AHBRlaunch,Leiden26}. We 
can furthermore use AHBR-QE to predict the structure and 
therefore also properties of so-called simple
solids \cite{Tran19,AHBRlaunch} because these are 
systems where there is only one relevant structural
parameter. For example, when there is just the one lattice constant $a$, we use AHBR-QE to compute cohesive energies at various $a$ values, establish an energy curve, and thus identify the lowest-energy form \cite{AHBRlaunch}.

The bottom panel of  Fig.\ \ref{fig:SchematicsVCrelax}
considers and compares present AHBR-QE and new 
AHBR-VASP strategies and results for completing 
structure optimization for graphite in terms of 
seeking the optimal height $c$ of the hexagonal
unit cell. It is a case where a simply structure 
optimization, completed in AHBR-QE (blue curve), may 
still work by making an approximation: We fix the 
in-plane lattice 
constant $a$ (at the value known from experiments) and compute the unit-cell energy 
as a function of a select set of $c$ values. We thus define a 
layer-binding
variation $E_{\rm bind}(c)$ upon subtracting the 
$c\to \infty$ energy and complete a fair structure
optimization by identifying the minimum $c_0$. In turn, with
AHBR-QE we may also estimate the layer binding using $E_{\rm bind} \approx E_{\rm bind}(c_0)$. In contrast, with the (here implemented) AHBR-VASP 
tool we leverage stress-based optimization to find the minimum allowing general $a$ and $c$ variations. We get the optimal structure and $E_{\rm bind}$ predictions (triangle) in one stress-based optimization study while avoiding
the approximation implied in the
curve-based AHBR-QE approach \cite{ziambaras03p064112}.

We note in passing that the AHBR is promising
for predicting structure in layered systems like 
graphite \cite{AHBRlaunch}; This ability is a requirement for 
understanding, in fact discovering, the stacking 
and three-dimensional assembly in C2N and 
many other COFs. For the graphite case,
comparison with experiment (dotted vertical
line) and with a random phase approximation (RPA) 
study \cite{OlTh2013} (dashed-dotted vertical lines,) shows that both the AHBR-QE and AHBR-VASP 
optimization strategy brings accuracy on the 
$c_0$ prediction. The same is also true regarding
the layer binding energy: Both predictions
fall within the range of numerical 
uncertainty associated with  
quantum Monte Carlo (QMC) studies \cite{Spanu09p196401,GaKiPa2014,MoDrFa2015,ShKiLe2017} (illustrated by the pair 
of horizontal dashed lines in the lower panel).
At the same time, alignment of results by AHBR-QE and 
AHBR-VASP optimization strategies is not automatic 
and that is not merely due to the simplification used in the AHBR-QE structure search: The AHBR-VASP uses the VASP PAW setups that are expected to
generally be more accurate than
what is possible using normconserving PPs (as we do in AHBR-QE).
  
Our interest in using AHBR for discovery motives moving to stress-based optimization even for relatively simple problems, e.g., those discussed in Fig.\ \ref{fig:SchematicsVCrelax}.
For example, the almost-rocksalt cases of the metal-monoxides, like MnO, suggests benefits of having stress-based optimization by AHBR. While proper rocksalt matter has exactly one lattice parameter ($a$), the `almost' is a key concern for AHBR modeling: We do not, \textit{a priori,} know how many and exactly which lattice parameters are needed to reflect relevant deformation. Similarly, the individual-sheet structure of C2N COF (or of
the earlier COF1 system), is well characterized as being consistent with a hexagonal cell, and being thus characterized by just one (in-plane) lattice constant $a$. Still, an AHBR-QE optimization may not suffice to discover the set of motifs that defines the
three-dimensional stacking;   The optimization may well alter the unit cell symmetry from hexagonal.
Generally, there exists a broad range of examples, 
general molecular crystals \cite{RanPRB16,Frostenson2024}, 
multiferroics like BiMnO$_3$, or hybrid organic-inorganic perovskites \cite{Hard2Soft,HybridPerov26}, 
where the extended-structure descriptions 
involve predicting most or all the six generic 
unit-cell parameters (lattice vectors and angles); The existing 
AHBR-QE approach will then require at least million RSH runs (each having atomic relaxations in fixed cells) and that is not practical.

\subsection{Analytical-hole range-separated hybrids: Design ideas 
and AHBR-HSE differences}

The design of the HSE \cite{EP98,HSE03,HSE06} begins with the popular PBE, but represents a robust generalization that permit
use in generalized KS DFT 
\cite{GKSstart,BeckeIII,Burke97,kronik2012,OTRSHalga,WiOhHa21,AHBRmRSH25}. HSE preserves key constraints of the PBE 
designs, for example,
spin scaling of exchange \cite{PerZun81,Thonhauser_2015:spin_signature} and charge conservation in the design of the so-called PBE XC hole  \cite{pewa86,pebuwa96}. As for crafting XC functionals with a GGA-type exchange (be it PBE, CX, B86R, or other), the first essential step towards a systematic RSH extension is crafting an exchange-hole model, here denoted $n_{\rm x}^{\rm DF}(\mathbf{r}; |\mathbf{r}-\mathbf{r'}|)$, that is short-range so that it is consistent also with descriptions of metals \cite{helujpc1971,lape77,EP98,HSE03,HSE06,HJS08,DefineAHCX,AHBRlaunch}.

The central underlying logic is securing full screening in the XC-energy contributions that arise at large separations between the electron and its associated XC and especially
the exchange hole \cite{gulu76,lape77,adawda}. This emphasis ensures seamless integration with the local-density approximation (LDA) in the 
limit of a homogeneous 
electron gas (HEG) so that HSE -- as well as the two 
recent RSH vdW-DFs -- permit studies of general systems, for example, avoiding spurious divergences at the Fermi level \cite{gulu76,lape77,lape80,lavo87,pewa92}.
A key argument in the design and use of the AHBR is that
the original HSE ideas can be directly ported, building instead 
on an analytical exchange-hole modeling for the B86R functional  \cite{DefineAHCX,AHBRlaunch,AHBRmRSH25}.

The designs of HSE and AHBR rely on crafting an oscillation-free representation of 
characteristic screened-exchange holes \cite{EP98},
denoted
$n_{\rm x}^{\rm DF}(\mathbf{r}; |\mathbf{r}-\mathbf{r}|)$. These holes reflect the nature of the GGA-type exchange descriptions \cite{gulu76,lape77,adawda,pewa86,pebuwa96,HJS08,DefineAHCX}, 
\begin{equation}
E_{\rm x}^{\rm DF} = \frac{1}{2} \, \int_\mathbf{r} \, n(\mathbf{r})
\int_\mathbf{r'} \, \frac{n_{\rm x}^{\rm DF}(\mathbf{r}; 
|\mathbf{r}-\mathbf{r'}|)} {|\mathbf{r}-\mathbf{r'}|} 
\label{eq:ExGGA}\, ,
\end{equation}
that are core parts of the underlying (regular, non-hybrid)
PBE or B86R functionals, respectively. Here and below we use the subscript on the integral symbols
to indicate a full-range spatial integration of the stated coordinate, e.g., $\mathbf{r'}$.
The exchange energy Eq.\ (\ref{eq:ExGGA}) is part of the energy  correction that must be made in DFT where the Hartree-energy evaluation represents an over-counting of the electron-electron interaction energy. One arrives at LDA
exchange, set by the local value of the Fermi wavevector $k_F(n)$, when respecting input that is defined in the HEG at various electron densities $n(\mathbf{r})$.

For XC functionals with a GGA type exchange such as PBE and B86R, one goes 
beyond the HEG by reflecting also the role in a local variation in the so-called scaled density gradient 
$s(\mathbf{r})=|\nabla n|/(2n(\mathbf{r})k_F(n(\mathbf{r}))$ \cite{pewa86}. This is done while still enforcing an asymptotic screening of exchange contributions in
Eq.\ (\ref{eq:ExGGA}) and seamless integration with the near-exact or
exact descriptions that are known for the HEG systems \cite{lu67,helujpc1971,Ceperley78,CepAld80,pewa92,pebuwa96,EP98,HSE03,HJS08,DefineAHCX}. 
One can therefore also rely on MBPT results on exchange in systems with gradient, along with an exact spin-scaling condition and further physics-based  constraints \cite{PerZun81,lavo87,pebuer96,pebuwa96,PBEsol,Dion,Thonhauser_2015:spin_signature}. 

For practical discussions one focuses the analysis on the scaled exchange hole 
\begin{equation}
J^{\rm DF}_{\rm x}(s(\mathbf{r}); y=|\mathbf{r}-\mathbf{r'}|k_F(n(\mathbf{r}))\equiv n^{\rm DF}_x(\mathbf{r}; y)/n(\mathbf{r})\,
.
\end{equation}
This hole model has a spatial dependence  \cite{HJS08,DefineAHCX,AHBRlaunch}
\begin{eqnarray}
    J_{\rm x}^{\rm DF}(s,y) & \equiv & {\cal{I}}(s,y) \exp(-s^2{\cal{H}}(s)y^2) 
    \nonumber \\
    & \sim & \exp(-[D+s^2{\cal{H}}(s)]y^2)
    \, , 
    \label{eq:ExHoleGauss}
\end{eqnarray}
dominated by a Gaussian form and 
all details of the $\cal{I}$ prefactor are implicitly set once the ${\cal{H}}(s)$ form is set.  The formal structure of the standard GGA-type exchange descriptions, 
\begin{eqnarray}
    E_{\rm x}^{\rm DF}
    & = &
    \int_\mathbf{r}
    \, n(\mathbf{r}) \varepsilon_x(n(\mathbf{r}))F_{x}^{\rm DF}(s(\mathbf{r})) \, ,
    \label{eq:SemilocalExformPBE}
    \\
    F_{x}^{\rm DF}(s) & = & -\frac{8}{9} \, \int_0^\infty
    \, y \, J_{x}^{\rm DF}(s,y) \, dy \, ,
\end{eqnarray}
emerges upon integrations \cite{HJS08,DefineAHCX}. Here $ \varepsilon_x(n(\mathbf{r}))$
denotes the LDA exchange energy per particle, set by $k_F(\mathbf{r})$.

Finally, to also define corresponding RSH extensions, like HSE and AHBR, these GGA-type exchange hole models $n_x^{\rm DF}$ 
are in turn used to isolate corresponding short-range (SR) exchange-functional contributions:
\begin{eqnarray}
E_{{\rm x, SR}}^{\rm DF}(\gamma)
& = &
\int_\mathbf{r} n(\mathbf{r}) \varepsilon_x(n(\mathbf{r})) F_{x,{\rm SR}}^{\rm DF}(k_F(n(\mathbf{r}), s(\mathbf{r})),
\\
F_{x,{\rm SR}}^{\rm DF}(k_F,s)
&= & 
    -\frac{8}{9} \int_0^\infty y J_{x}^{\rm DF}(s,y) \erfc{(\gamma y/k_F)} 
    \, dy.
\label{eq:ExSR}
\end{eqnarray}
Here `$\erfc$' denotes error-function compliment, ensuring that
contributions at large $|\mathbf{r}-\mathbf{r'}|$ values are suppressed.
Meanwhile, we can from any given approximation of the DFT orbitals evaluate a correspondingly defined SR component of Fock exchange $E_{\rm x,SR}^{\rm Fo}(\gamma)$, using Eq.\ (\ref{eq:ExGGA}) but replacing 
$n_{\rm x}^{\rm PBE}$ by the Fock-exchange hole
defined by the one-particle 
density-matrix approximation,
as summarized in Ref.\ \onlinecite{DefineAHCX}. 

Combining or merging these two descriptions of the exchange hole,
we arrive at the HSE or AHBR exchange-energy description 
\begin{equation}
E_{\rm x}^{\rm RSH-DF}(\alpha,\gamma) = 
\alpha [E_{\rm x,SR}^{\rm Fo}(\gamma)
- E_{\rm x,SR}^{\rm DF}(\gamma)]
+E_{\rm x}^{\rm DF} \, .
\end{equation}
The hybrid-parameter $\alpha$ denotes the extent of SR 
Fock-exchange mixing; The asymptotic electron-hole coupling in 
Eq.\ (\ref{eq:ExGGA}) is assumed to be completely set by the $E_x^{\rm DF}$
form. The resulting RSH XCs for generalized DFT therefore 
also reflects an inverse length scale $\gamma$, controlling the 
rollover between the SR coupling and the
asymptotic one \cite{HSE03,OTRSHalga}. The popular PBE-based RSH `HSE06' (which we will also use
for a few comparison studies) results upon making the standard settings $\alpha_0=0.25$
and $\gamma_0= 0.2$ {\AA}$^{-1}$ (0.106 inverse Bohr). For AHBR, we deliberately keep the settings of the $\alpha$ and $\gamma$ values fixed at those values; This holds for the default AHBR-QE version
and for the new AHBR-VASP implementation that is defined, discussed, and illustrated below.

\begin{figure*}
\centering
%
%
%
%
%
\includegraphics[width=0.91\linewidth]{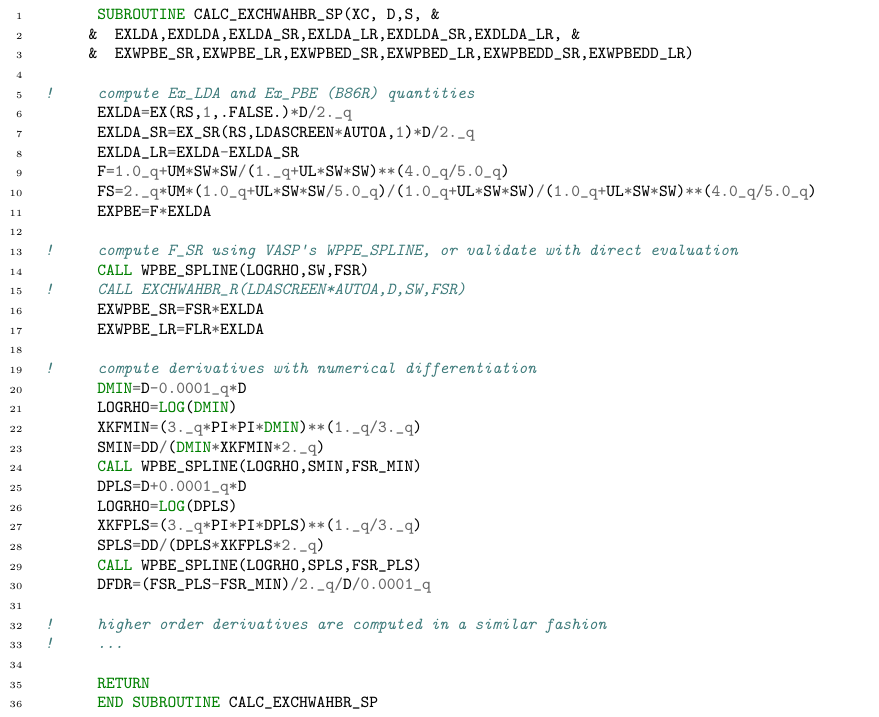}
\caption{\label{fig:minted}
  Synopsis of the functional-lookup subroutine excerpted from the implementation for VASP (version 6.5.1) that
  makes available fast extraction
  of the SR component, $F_{\mathrm{x,SR}}^{\rm b86r}$,
  of the exchange-enhancement factor in vdW-DF2-b86r.
  The AHBR-VASP functional energies and derivatives are made 
  available via the VASP spline-look table, in `WPBE\_SPLINE' subroutines. When we specify a AHBR-VASP study via the VASP 
  \texttt{INCAR} file, this spline table have first been initialized by the results of a set of calls to subroutine `CALC\_EXCHWAHBR\_R'.
  That spline-table initialization (not shown) provides the 
   underlying information on the $F_{\mathrm{x,SR}}^{\rm b86r}(\gamma;n,s)$ variation with density $n$ and density gradient 
   $s$, at the default value of $\gamma$ (or as controlled by
   the \texttt{INCAR} file, if desired).}
\end{figure*}

The main conceptual difference between the AHBR and the HSE arises with the specification of the correlation functional $E_c^{\rm DF}$. In both cases this description is unchanged from that of the underlying functionals (B86R and PBE repectively).
However the vdW-DFs, and AHBR in particular, stands out by having a truly nonlocal correlation component that captures vdW interaction effects while enforcing current conservation \cite{hybesc14,JPCMreview}. It does so by counting the integrated energy shift that arises by the electrodynamic coupling \cite{jerry65,ryluladi00,JPCMreview} of a set of collective excitations (plasmons) \cite{lape80,lavo87,ryluladi00,rydberg03p126402,hybesc14}. This starting point 
reflects, in turn, the zero-point electron dynamics as captured 
in a simpler, so-called internal, GGA-type functional \cite{lee10p081101,hybesc14,Berland_2015:van_waals,JPCMreview}
\begin{equation}
E_{\rm xc}^{\rm in} = \int_0^\infty 
\frac{du}{2\pi}\, \hbox{Tr}
\{ 
\ln(\epsilon(\omega))\} - E_{\rm self}\, .
\end{equation}
Here the trace represents integration over the spatial coordinates and $E_{\rm self}$ represents the standard removal of a divergence arising in the underlying MBPT response formulation \cite{gulu76,lape77,Dion,thonhauser,JPCMreview}. Noting that the plasmon poles of the internal functional represent virtual current fluctuations, however, we must also leverage the longitudinal projections implied in the continuity equation \cite{rydbergthesis,hybesc14}. We thereby upgrade the dielectric-function approximation $\epsilon \to \kappa_{\rm ACF}(\mathbf{r},\mathbf{r'};\omega)$, as discussed, for example, in Ref.\ \onlinecite{hybesc14}. This leads to truly nonlocal correlation where the beyond-LDA component can be expressed from the dielectric-function differences \cite{Berland_2015:van_waals,JPCMreview}
\begin{equation}
    E_{\rm c}^{\rm nl} \equiv  
    \int_0^\infty 
\frac{du}{2\pi}\, \hbox{Tr}\{ 
\kappa_{\rm ACF}(\epsilon(\omega))\} 
    - \int_0^\infty 
\frac{du}{2\pi}\, \hbox{Tr}\{ 
\ln(\epsilon(\omega))\} \, .
\label{eq:Ecnlformal)}
\end{equation}
The upgrade $\epsilon \to \kappa_{\rm ACF}$ of the electrodynamics description is set by enforcing current conservation \cite{hybesc14,JPCMreview}.

The popular, computationally efficient (general-geometry) vdW-DFs are defined by stopping the expansion in $\ln(\epsilon(\omega))$ after including all second-order terms. Upon multiple formal integrations \cite{Dion,rydbergthesis,dionthesis} one arrives at a simple-two-coordinate formulation
\begin{equation}
E_{\rm c}^{\rm nl} \approx
\frac{1}{2}\, \int \int n(\mathbf{r}) \Phi_{\rm c}^{\rm nl}(\mathbf{r},\mathbf{r'})\, n(\mathbf{r'})
\end{equation}
that is defined by a universal nonlocal-correlation kernel $\Phi_{\rm c}^{\rm nl}$ \cite{Dion,thonhauser}. This kernel depends on the electron density and scaled-density gradient at 
positions $\mathbf{r}$ and $\mathbf{r'}$ but remains 
density explicit. This gives a fast evaluation of such nonlocal correlation effects and this holds true also for AHBR studies. The consistent inclusion of all second-order $\ln(\epsilon)$ terms means that the vdW-DFs have seamless integration for LDA in the HEG limit
\cite{Dion,JPCMreview}.

This paper includes a set of AHBR-PAW-VASP and AHBR-PP-QE calculations to illustrate
the importance of stress-based optimization. The AHBR-QE calculations are standard, part of the QE
release since about 2022. This paper also includes HSE-PAW-VASP calculations for comparison of the description of the almost-rocksalt metal-monoxide cases; HSE itself is not set up to predict the structure of layered systems. Table \ref{tab:INCARset} presents an overview of differences among VASP-input parameters relevant for 
using AHBR-VASP instead of HSE06-VASP for stress-based structure optimization in
our here-documented code work.

\subsection{Subroutine extension and call}

The implementation of any functional requires the access to subroutines for fast, accurate evaluation of the XC energy  $E_{xc}$. We also need the
derivatives of these components  with respect to 
$n$ and $\nabla{n}$ or the scaled density gradient $s$. This is necessary because they are to be used with the self-consistency cycles of the KS solution scheme. VASP already
offers simple access to, for example, a choice of running DFT with the rev-vdW-DF2 (i.e., the vdW-DF2-b86r) XC functional; Those parts represent no challenge for the AHBR implementation work.

For AHBR, as for HSE, usage in DFT we must also have access to subroutines for fast determination of SR Fock
exchange term $E_{\rm x,SR}^{\rm Fo}$. The same is true of relevant SR component of the associated GGA exchange, $E_{\rm x,SR}^{\rm PBE}$ and $E_{\rm x,SR}^{\rm b86r}$. Since there have long been support for HSE, the code parts for $E_{\rm x,SR}^{\rm Fo}$ and $E_{\rm x,SR}^{\rm PBE}$ are readily available. 

In the AHBR-QE implementation we provide
an evaluation of $E_{\rm x,SR}^{\rm b86r}$ and its derivatives
by there adding a subroutine `ahsx' (based on the HJS hole formulation \cite{HJS08,DefineAHCX,AHBRlaunch}) adapting an original QE subroutine
for using the EP-hole description \cite{EP98,HSE03}). Here we complete the AHBR-VASP implementation in three steps:
 
\begin{enumerate}
    \item Provide a subroutine 
    \verb|CALC_EXCHWAHBR_R| (corresponding to the QE `ahsx' subroutine \cite{DefineAHCX,AHBRlaunch}) 
    for the computation of the SR component of the exchange-enhancement  factor $F_{\mathrm{x,SR}}^{\rm b86r}(\gamma;n,s)$;
    \item Compute a spline table of the $F_{x,\mathrm{SR}}^{\rm b86r}$ 
    and hence $E_{\mathrm{x,SR}}^{\rm b86r}$ variation with $n$ and $s$; 
    \item Write a wrapper \verb|CALC_EXCHWAHBR_SP| subroutine
    which on demand helps return the value and derivatives of the RSH vdW-DF $E_{\rm xc}$ via lookup from the \verb|WPBE_SPLINE| table.
\end{enumerate}

For the first step we note that VASP authors have 
introduced the so-called HSEsol 
\cite{HJS08,HSEsol} as a consistently-defined 
RSH-GGA that is documented as a viable alternative to HSE, at least for bulk systems.   
The HSEsol-VASP is crafted by using both the HJS  
exchange-hole framework \cite{HJS08} and the specific 
PBEsol-exchange parametrization provided 
in Ref.\ \onlinecite{HJS08}. HSEsol is promising
for bulk-metal cases where PBEsol is expected to perform 
better than PBE \cite{HSEsol}; In connection with launching AHBR-QE \cite{AHBRlaunch}, we used the `ahsx' subroutine to also 
add an `HSEsol' option (by setting the QE `\verb|input_DFT|' 
parameter). For the present AHBR-VASP focus, we note 
that the VASP source file `wpbe.F' already includes 
a HJS08-based subroutine \verb|CALC_EXCHWPBEsol_R|. 
We adapt that VASP subroutine form, while adjusting the parameters reflecting the nature of the exchange-hole modeling \cite{HJS08,DefineAHCX} to the values relevant for B86R \cite{AHBRlaunch}. 

For the second step, we simply had to insert extra options 
for \verb|WPBE_SPLINE| subroutines, again in `wpbe.F',
to realize initialization by 
calls to \verb|CALC_EXCHWAHBR_R|.

Figure \ref{fig:minted} summarizes the core part of the
XC-functional wrapper that we add to the VASP source code
`ggalib.F' so that the rest of the VASP DFT-solver framework
can easily access relevant $F_{\rm x,SR}^{\rm b86r}$ and $E_{\rm xc}$ information. This is done, as in the HSEsol case, by lookup calls to \verb|WPBE_SPLINE|, that in turn is filled at DFT upstart with info that 
\verb|CALC_EXCHAHBR_R| provides. The coding is shown in Fig.\ \ref{fig:minted} as it appears in VASP version 6.5.1; There are essentially no differences from the coding used for VASP 5.4.1 or VASP 6.2.0, namely,
the version used for almost all of our test and analysis 
(with exceptions explicitly noted when relevant for our 
documentation and demonstrator studies.)

Figure \ref{fig:minted} also shows a code line (\#15, commented out)
that can be used to test the precision of using the spline-look-up table
for XC calls. In practice, this AHBR implementation 
testing was done by comparing accuracy in AHBR-VASP when instead we used
direct \verb|CALC_EXCHWAHBR_R| calls inside this subroutine, 
essentially by reversing the commenting used for lines \#14 and \#15 shown in Fig.\ \ref{fig:minted} (and everywhere 
else \verb|WPBE_SPINE| is called).  
We found no observable impact of using the spline-lookup-table
approach for AHBR-VASP studies. 

More generally, our AHBR-VASP implementation has also been 
tested for different combinations of CPU architectures, 
compilers, and parallelization setups. By the COF 
demonstrator study (that uses VASP version 6.5.1) our AHBR-PAW-VASP
implementation has furthermore been tested for use with GPU architectures.

\begin{figure}
    \centering
    \includegraphics[width=1.\linewidth]{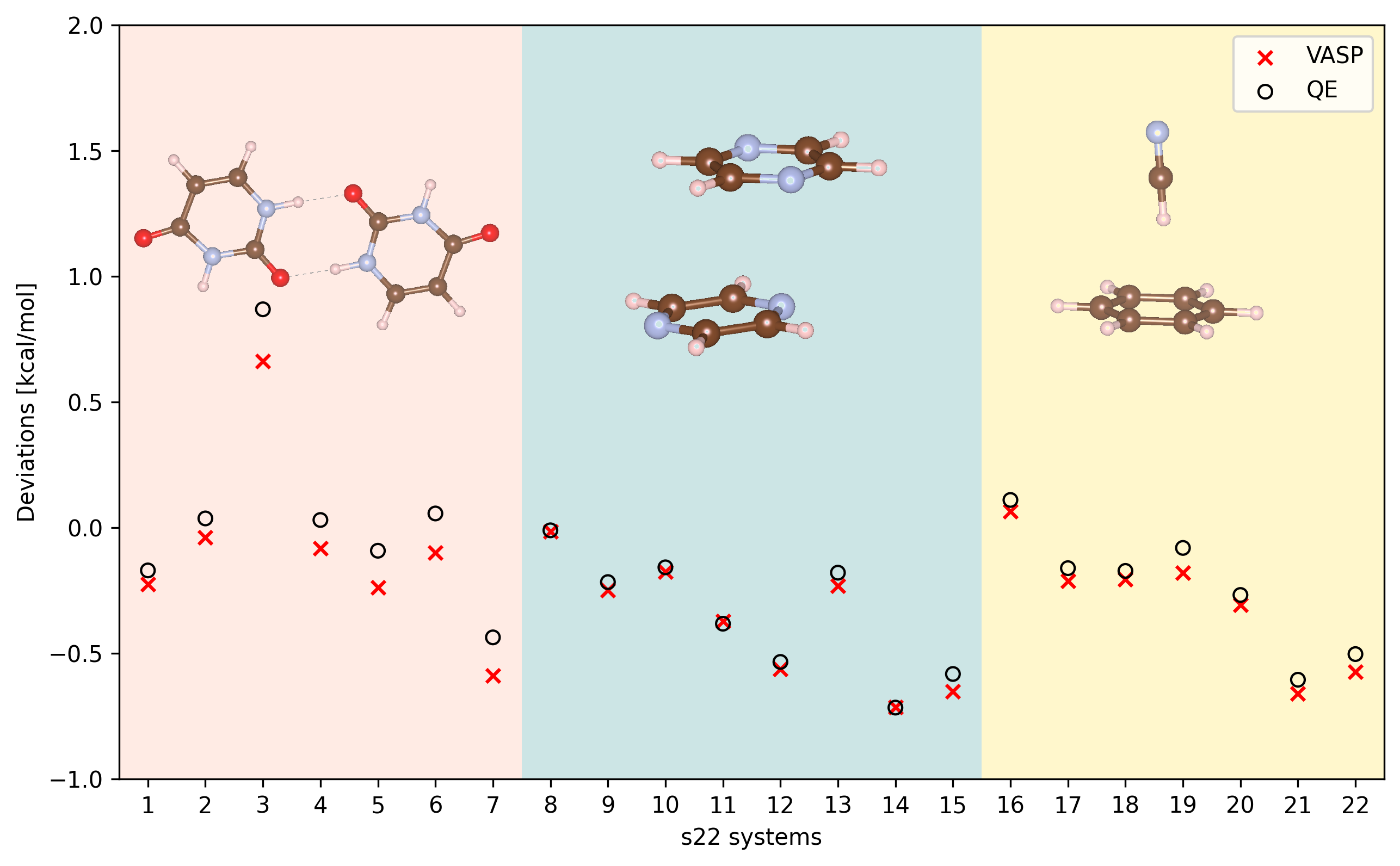}
    \caption{Validation of AHBR-VASP implementation as tested on the 
    set of hydrogen-bonded (left), dispersions-dominated (middle), and mixed
    systems (right) in the S22 benchmark set.
    For each such system `\#', we  report deviations, $E_{\rm bind}^{\rm QE,\#}-E_{\rm bind}^{\rm S22,\#}
    $ and $E_{\rm bind}^{\rm VASP,\#}-E_{\rm bind}^{\rm S22,\#}$
    (in kcal/mol), from S22 binding-energy reference data, $E_{\rm bind}^{\rm S22,\#}$ 
    by our AHBR-VASP (AHBR-QE) characterizations $E_{\rm bind}^{\rm VASP,\#}$ ($E_{\rm bind}^{\rm QE,\#}$). Benchmarks 1-7 concern  hydrogen-bonded complexes, exemplified by 
    benchmark 5 (the uracil dimer). Benchmarks 8-15 concern dispersion-dominated complexes, exemplified by benchmark 12 (the pyrazine dimer). Benchmarks 16-22 concerns mixed-interaction complexes, exemplified by benchmark 19 (the benzene-HCN complex).
    }
    \label{fig:S22Converge}
\end{figure}

\section{Computational details}

For AHBR-VASP (and corresponding HSE) calculations we use the set of PAW setup denoted PBE-6.4. Within and for any specific atom, we systematically pick the PAW version that comes with the largest stated requirement on the ENCUT setting, while avoiding 
versions marked `GW'. For the AHBR-(PAW-)VASP studies, we systematically pick an actual ENCUT value (even in convergence tests) that is significantly larger than all of those recommendations stated in the PAW setups being used. All production runs for simple solids 
and for the anti-ferromagnetic metal monoxides are characterized at ENCUT=1600; We find that a smaller ENCUT would often be sufficient here.
However, our choice generally makes it easy to find the right spin state of atoms, and thus of the cohesive energies $E_{\rm coh}$. 

A smaller ENCUT=1200 eV together with a $4\times 4\times 4$ $k$-point
sampling are used for AHBR-VASP predictions of stacking order 
in the C2N covalent-organic framework (COF) system \cite{C2Nstart2015,LeeGapNanoPore2025},
contrasting CX and AHBR results; In the case of the latter, the sampling 
of $k$-point differences (the $q$ point sampling) is set as a $2\times 2\times 2$ grid. The lower $k$-point sampling is motivated by the fact that the unit-cell for a two-layer COF-sheet stacking is substantially larger than for the simple-bulk or monoxide cases.
For computations of the 2CN band structure details, including an
exploration of effects  doped
C2N systems, we increased the sampling over $q$ points
to a $4\times4\times4$ grid in 
frozen self-consistent calculations.

All system and atom energies (in AHBR-VASP or AHBR-QE) are converged to at least a micro-eV. In our AHBR-VASP studies the setting (in the INCAR file)  ``EDIFF = 10$^{-6}$'' is used generally but ``EDIFF = 10$^{-8}$'' is used for structural relaxations in the COF systems. This caution is taken because we find that the three-dimensional stacking in COF systems is often set by a subtle competition between steric hindrance, electrostatic, and vdW forces.
For the many variable-cell AHBR-VASP studies, we use an 
``EDIFFG=10$^{-3}$'' setting  for accuracy in predictions of unit-cells 
and atom configurations.

For layered systems (Fig.\ 1), and for comparison and implementation testing, we also provide (or, in a few cases, repeat from
Ref.\ \onlinecite{AHBRlaunch}) 
AHBR-QE calculations. Here, for 
simple-bulk cases we provide test of convergence with regards to sampling of $k$-points as well as of $k$-point-differences (so-called $q$-points), as summarized in the appendix. We use the electron-rich ONCV-SG15 PPs at a 160/640 Ry wavefunction-/density-energy cutoff, to facilitate simpler comparison with the original AHBR paper \cite{AHBRlaunch}. This is also useful for simplifying comparisons
with a recent
papers documenting usefulness 
of this RSH vdW-DF for CO site preference and H$_2$ dissociate adsorption on metal surfaces \cite{AHBRlaunch,Leiden26}.

\begin{figure*}
    \centering
    \includegraphics[width=0.95\linewidth]
    {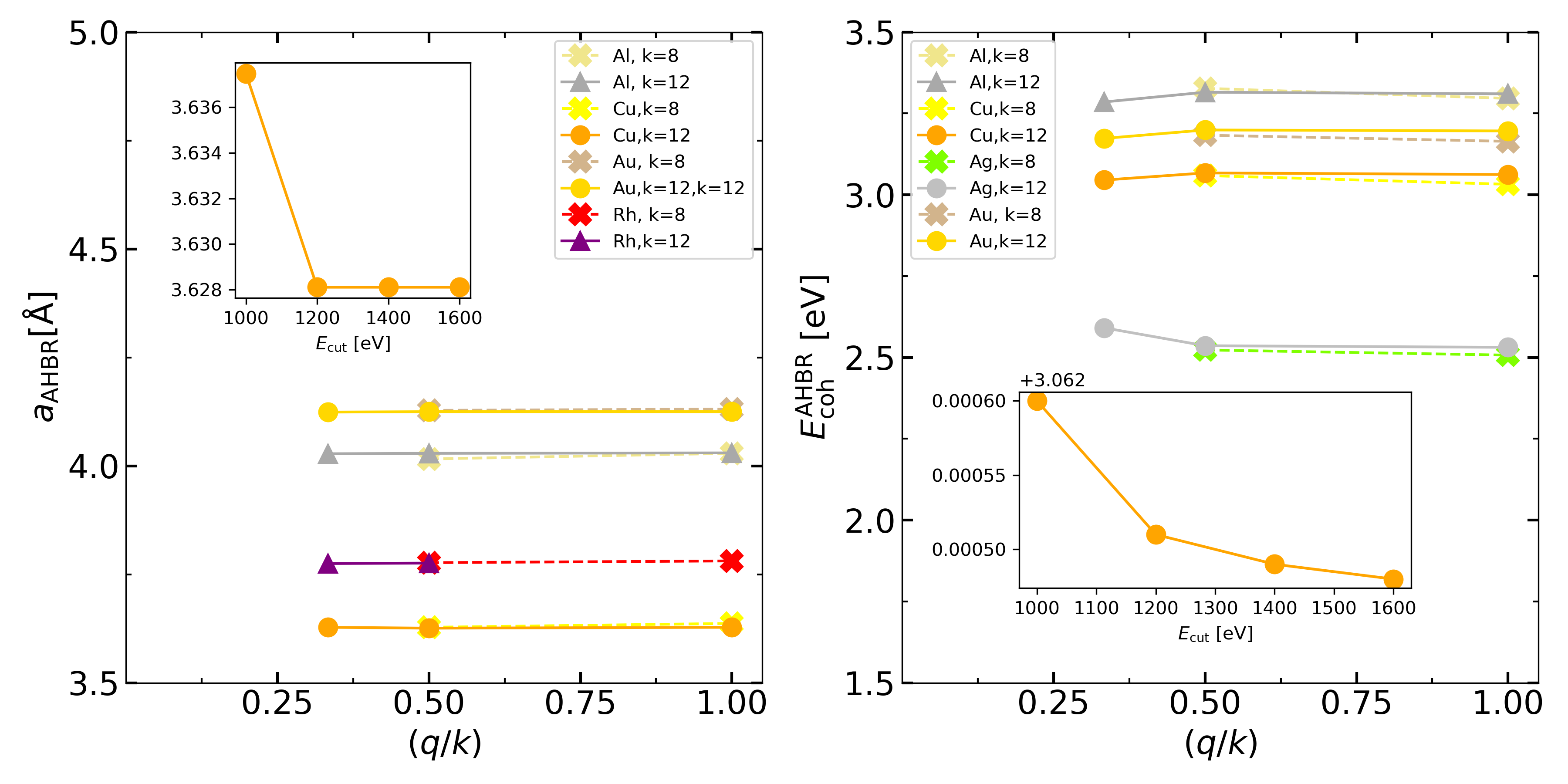}
    \caption{Convergence with respect to $k$ and $k$-point differences $q$ of the AHBR-VASP predictions for metals.
    Insert shows convergence with ENCUT for Cu at $k=q=12$.
    }
    \label{fig:MetalConverge}
\end{figure*}

Our choice of a systematic large ENCUT values is further motivated by our use of reference bulk systems for tests of the variable-cell
optimization path that our AHBR-VASP implementation enables. A bulk-system benchmark 
study \cite{Tran17,Tran19}
provides reference values 
for the (single) lattice 
constants $a$ and cohesive energies $E_{\rm coh}$. These
are based on experiments but
back-corrected for (zero-point and thermal) vibrational 
effects. We contrast the AHBR-VASP and AHBR-QE performance
on a subset of the reference data \cite{,Tran16,Tran19},
noting that the AHBR-QE optimization path remains feasible 
here. The comparison of deviations among the two strategies 
in turn tests our implementation of stress-based AHBR-VASP 
structure optimization, as long as we ensure full convergence 
in both AHBR codings, e.g., on ENCUT for AHBR-VASP studies.

A more important sensitivity
to convergence, however, arises 
by the $k$ and $q$ sampling.
We therefore summarize our documentation (fully detailed again in the SI document) of such AHBR-VASP convergence finding permitting us to state
that $(k=12,q=6)$ samplings 
generally suffice for single-atom metallic test cases; For the multiple-atom semiconductor and insulator cases, we can generally work with a $(k=8,q=8)$ sampling. At the same time, our AHBR-VASP predictions of metal-monoxide
systems (in the anti-ferromagnetic, type II, rocksalt 
phase) illustrate that convergence tests are important, at least with regards to the $q$ sampling that controls accuracy in the AHBR Fock-exchange evaluation.

\section{Results and discussion}

\subsection{Validation by molecules: S22 } 

Figure \ref{fig:S22Converge} summarizes our first test of the AHBR-VASP implementation, namely by comparison of performance on molecular-energy differences, as asserted via the S22 benchmark set \cite{gmtkn55}. 
This set comprises 22 molecular binding-energy cases defined via reference geometries and
corresponding reference
values obtained via high-accuracy symmetry-adapted
coupled-cluster CCSD(T) studies. 
The benchmark set considers seven cases of problems where vdW binding
dominates, seven reflecting hydrogen bonds, as well as eight mixed-binding cases. The benchmark 
set thus provides 22 total-energy differences, $E_{\rm Bind,\#}$, between the product (assembly of multiple molecules) state and reactants (isolated molecules)
with which we can compare AHBR
performance \cite{AHBRlaunch}.

The figure contrasts our set of 
22 AHBR-VASP (crosses) and 
22 AHBR-QE (circles) binding-energy results, in terms of their deviation (measured in kcal/mol) from the benchmark values. 
The data obtained in the underlying set of binding-energy descriptions are summarized in appendix A. We note in passing that we use the same large unit cell as in Ref.\ \onlinecite{AHBRlaunch}. However,
unlike in those AHBR-QE studies, we did not use the option for 
an electrostatic decoupling \cite{Makov} between the molecules in the
repeated unit-cell representation used in planewave codes. 
This is done to avoid any
potential offset that simply reflects differences in 
QE/VASP strategies for electrostatic decoupling between repeated images.

We find that the AHBR performs well (in both codes) on the S22 benchmark set, as also expected from the analysis on the role of the unit-cell size that we presented in Refs.\ \onlinecite{DefineAHCX,AHBRlaunch}.
There are small differences but these are in part expected because use of AHBR-VASP also means using the PAW setups instead of PPs.

\begin{figure*}
    \centering
    \includegraphics[width=0.95\linewidth]
    {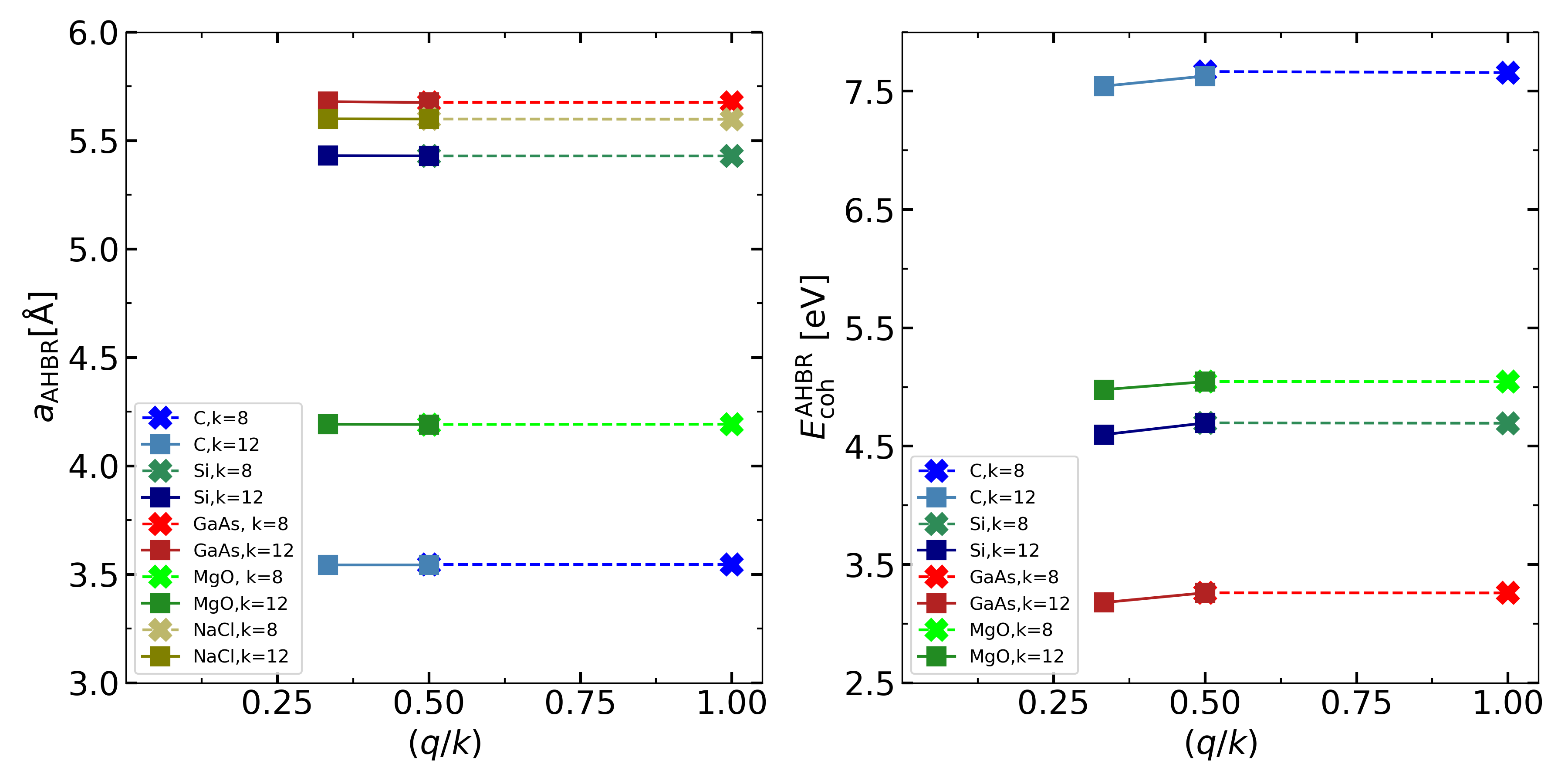}
    \caption{Convergence with respect to $k$ (and $k$-point differences $q$) of the AHBR-VASP predictions for semiconductor/insulator systems.
    }
    \label{fig:InsulatorConverge}
\end{figure*}

\subsection{Validation by simple-bulk reference systems}

We next turn to consider bulk examples taken from the so-called Solid-44 benchmark set \cite{Tran16,Tran19}, i.e., simple bulk systems
where there is only one relevant lattice constant and where the experimental structure is known from experiments. The bulk cohesive
energy $E_{\rm coh}$ is defined as the difference (per unit-cell atom) 
in the energy of the assembled bulk and that of the set of constituent atoms, when these are considered in isolation. The $E_{\rm coh}$ values
together with corresponding 
optimal lattice-constants values, denoted $a$, are   
back-corrected for 
vibrational zero-point energy and thermal
expansion effects in the Solid-44
benchmark set \cite{Tran16,Gharaee2017,Tran19}. 
Accordingly, one can directly compare Solid-44 benchmark values with DFT results based on, for example, AHBR. 
 
The use of (metal and semiconductor) cases from the Solid-44 set permits us an experiment-guided test of consistency among AHBR-VASP and AHBR-QE structure optimization results, essentially as illustrated in Fig.\ \ref{fig:SchematicsVCrelax}. We can test and validate our present implementation work by noting that for cases with simple-bulk structure, there should be no AHBR-VASP and AHBR-QE differences other than those reflecting that the former uses PAW setups while the latter uses norm-conserving PPs. That is, as long we push all calculations to full convergence, there should be at least some cases where there are no differences in the AHBR-VASP and AHBR-QE results. Also, given the promise of now enabling use of PAW setups, 
the deviations of AHBR-PAW-VASP results from
the bulk-system reference data \cite{Tran19}
should generally be smaller than those of AHBR-PP-QE results.

To deliver such an implementation test in practice, we first
document that we do indeed ensure convergence of both the AHBR-VASP and AHBR-QE results in their $a$ and $E_{\rm coh}$ predictions for simple-bulk cases, next.

\subsubsection{Convergence with regards to energy cutoff}

\begin{figure*}
    \centering
    \includegraphics[width=0.95\linewidth]
{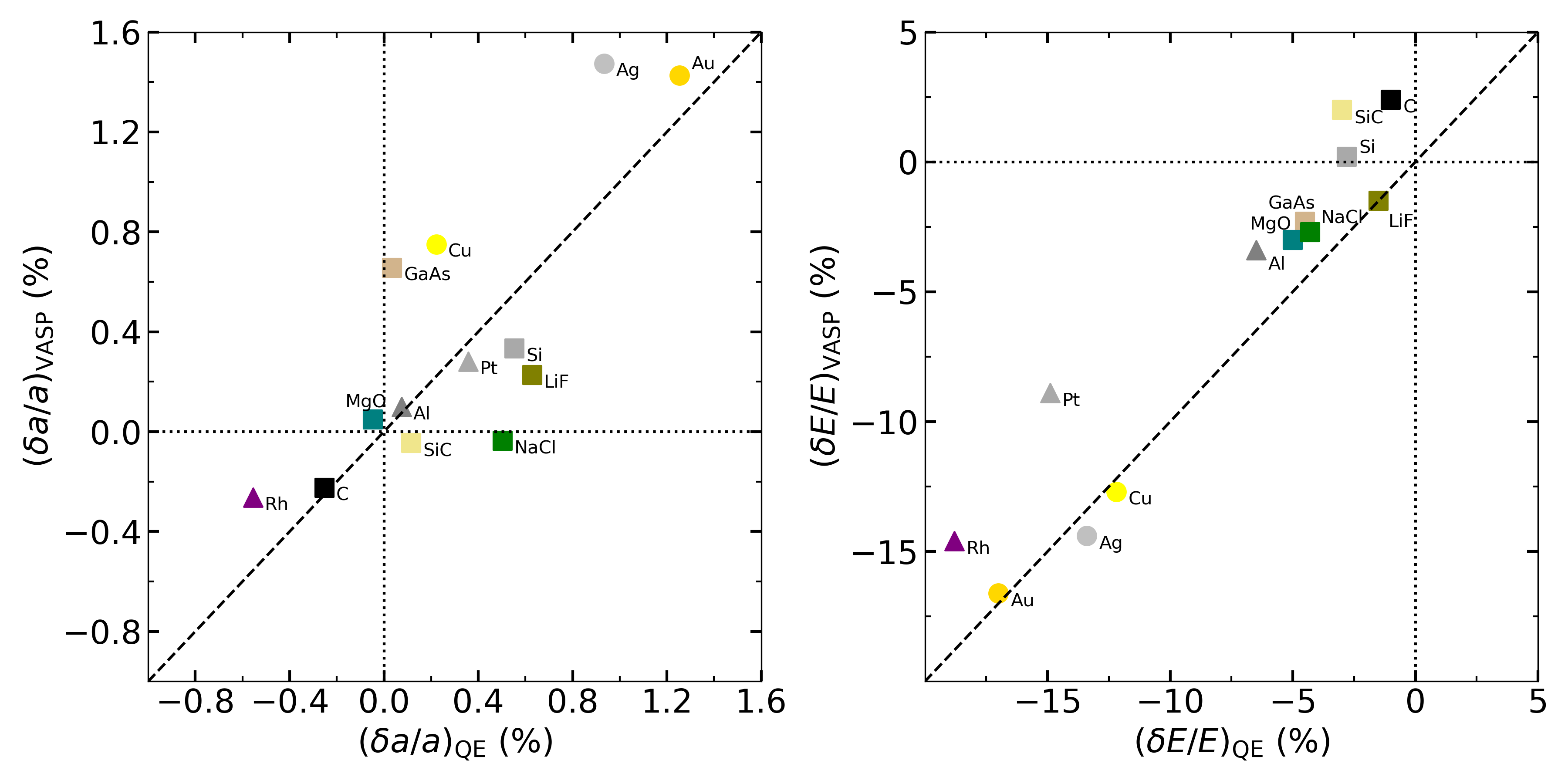}
    \caption{Accuracy of AHBR-VASP versus AHBR-QE predictions on subset of metallic and semiconductor/insulator cases selected from 
    a bulk-system benchmark set \cite{Tran19}. 
    }
    \label{fig:CROSScheckConverge}
\end{figure*}

Appendix B summarizes our tests of convergence with regards to the (wavefunction) energy cutoff 
(`ENCUT') and $k$-point samplings that we have completed for AHBR-VASP studies of metals, semiconductors, and metal-monoxides. Convergence is mainly asserted in terms of results for $E_{\rm coh}$ and for the optimal value of the main lattice constants $a$ in each material. With the 
focus on also metal monoxides we illustrate that AHBR-VASP brings 
an option to now complete  stress-bases variable cell optimization in systems with lower symmetry, i.e., existence of slight distortion (differing lattice constant $a'$) in one direction,  $a'\neq a$. For these metal-monoxide problems we also assert convergence in terms of the $a'/a$
ratio and the existence of an  
anti-ferromagnetic order .

The pair of inserts in panels Fig.\ \ref{fig:MetalConverge},
concerning the AHBR-VASP description of Cu structure and cohesion, respectively, show that the use of ENCUT=1600 eV suffices to provide a high convergence in VASP studies. These inserts show that such a high ENCUT value is required to 
get AHBR-VASP precision, beyond the three-digit focus that we reflect in
the tables of appendix B.

In summary, for the AHBR-VASP validation purposes, we find it motivated to systematically work with a VASP ENCUT value of 1600 Ry. The choice of such a high ENCUT value is also motivated by the fact that we shall be comparing with AHBR-QE studies provided at 160 Ry (that is, an even higher energy cutoff), and we want to minimize the risk of offsetting the comparison of our two AHBR implementations. As expected by the implied use of PAW setup and as suggested in part by Appendix B, it will likely often be enough to proceed with AHBR-VASP with a smaller ENCUT value. Nevertheless, when seeking to compute the NiO cohesive energy we do not get the right spin-polarization for the Ni atom at lower ENCUT values.
Also when checking corresponding HSE06-VASP descriptions of NiO, we found 
some dependence in the predictions for $a'/a$ values with the ENCUT choice.

\subsubsection{Convergence with regards to $k$-point sampling}

Figures \ref{fig:MetalConverge} 
and \ref{fig:InsulatorConverge} document
that we have secured convergence of the AHBR-VASP results for a set of metal and semiconductor cases from Solid-44 benchmark set \cite{Tran19}. More broadly, the figures summarize our convergence testing, Appendix B, with regards to the $k$-point sampling as well as with regards to $q$, that is, the set of $k$-point differences we include in the evaluation of the SR Fock-exchange component of the AHBR functional.

Figure \ref{fig:MetalConverge} shows that (given our choice of large
energy cutoffs) we can treat metal systems at $k=12$ while 
reducing the $q$ values to half or even a third of that. Meanwhile, 
if we were exclusively focused on the structure, it would 
often be sufficient to even limit the $k$ point sampling
to $k=8$, at least if we also chose no reduction in the $q$ point sampling ($q=k=8$); Appendix B shows that a similar set of criteria applies also for AHBR-QE descriptions. However, the right panel of the figure 
(concerning descriptions of $E_{\rm coh}$ values) also makes it 
clear that we can only get a precise comparison of AHBR-VASP and AHBR-QE  when going to descriptions based on a larger (more computationally
involved) set of $k,q$ choices.

Figure \ref{fig:InsulatorConverge} presents a corresponding analysis 
for semiconductor and insulating bulk cases
picked from the Solid-44 benchmark set \cite{Tran19}; Again the appendix presents a detailed listing of both AHBR-VASP and AHBR-QE results. The figure shows  
that at increasingly accurate picks of the $k$-$q$ combination, we improve the AHBR convergence on both lattice constants and cohesive energies.
We find that for a comparison of AHBR-VASP and AHBR-QE descriptions
in such non-conducting cases we can proceed at $k=q=8$ samplings.

\begin{table}[h]
\caption{\label{tab:MOsummary}
VASP variable-cell optimization of basic lattice constants $a$ and for the adjusted constant $a'$ (corresponding to the direction representing the anti-ferro-magnetic order `2') for metal-monoxides as described in AHBR (and HSE06 when explicitly noted).  The system tags, e.g., MnO-8-4, reflect the number of $k$ and $q$ points used in our 4-atom unit cell; Additional labels "(c)" identify variable-cell AHBR-VASP calculations in which we enforce cubic symmetry under relaxations. Experimental lattice constant values collected in Ref.\ \cite{Tran19}. 
The exclamation mark warns that we were not able to get a finite magnetization on the Ni-atom description (and thereby complete the $E_{\rm coh}$ prediction) using HSE06, at ENCUT=1600 eV, to facilitate a comparison.}
\begin{tabular}{l|ccccc}
\hline
\hline
Rocksalt system & $a$ [{\AA}] & $a'/a$ & $\mu$ [Bohr] & $E_{\rm coh}$ [eV] & $\Delta_{{\rm g}}$ [eV]\\
\hline
MnO-8-4  & 4.441  & 0.985  & 4.350 &  4.663 & - \\
MnO-8-4(c)  & 4.410  & (1.0)  & 4.352 &  4.662 & - \\
MnO-8-8  & 4.442  & 0.985  & 4.478 &  4.663 & 4.0 \\
MnO Exp.  &  4.445 &  -  &  4.58-4.79 & - & 3.9 \\
MnO-8-8 (HSE)  & 4.459  & 0.985  & 4.370 &  4.508 & 3.9 \\
\hline
NiO-8-4  & 4.166  & 0.998  & 1.664 &  4.541 & - \\
NiO-8-4(c)  & 4.163  & (1.0)  & 1.664 &  4.543 & - \\
NiO-8-8  & 4.167  & 0.998 & 1.664 &  4.543 & 4.5 \\
NiO Exp.  &  4.171 &  - & 1.64-1.90 & - & 4.0/4.3 \\
NiO-8-8 (HSE) &  4.181 &  0.998 & 1.684 & ! & 4.6 \\
\hline
FeO-8-4  & 4.280  & 1.008  & 3.485 &  4.679 & - \\
FeO-8-4(c)  & 4.298  &  (1.0) & 3.484 &  4.662 & - \\
FeO-8-8 & 4.342  & 0.979  & 3.478 &  4.673 & 2.2 \\
FeO Exp.  &  4.334 &  -  &  3.32-4.2 & - & 2.4 \\
\hline
\hline
\end{tabular}
\end{table}

\begin{figure*}
    \centering
    \includegraphics[width=0.975\columnwidth]
    {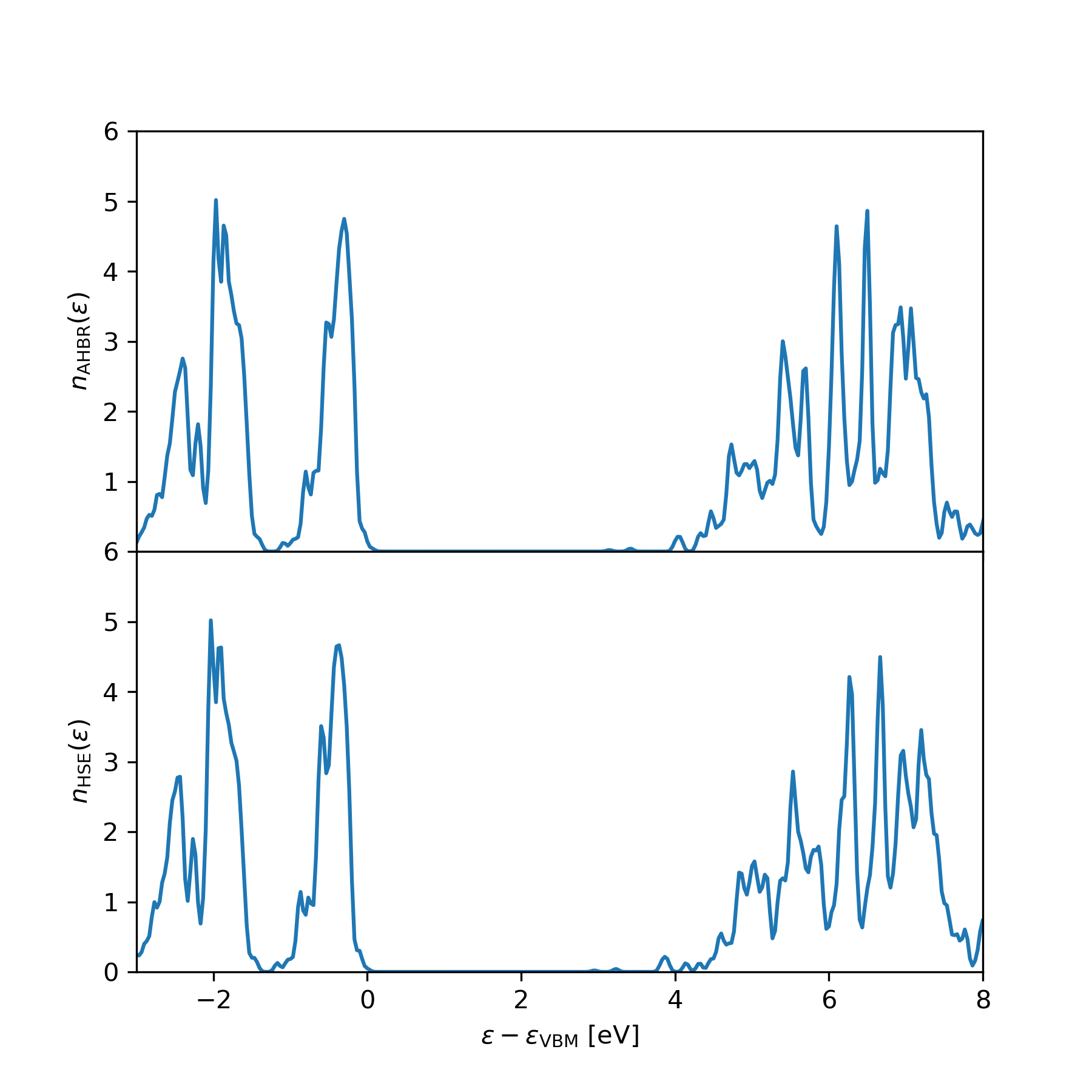}
\includegraphics[width=0.975\columnwidth]
{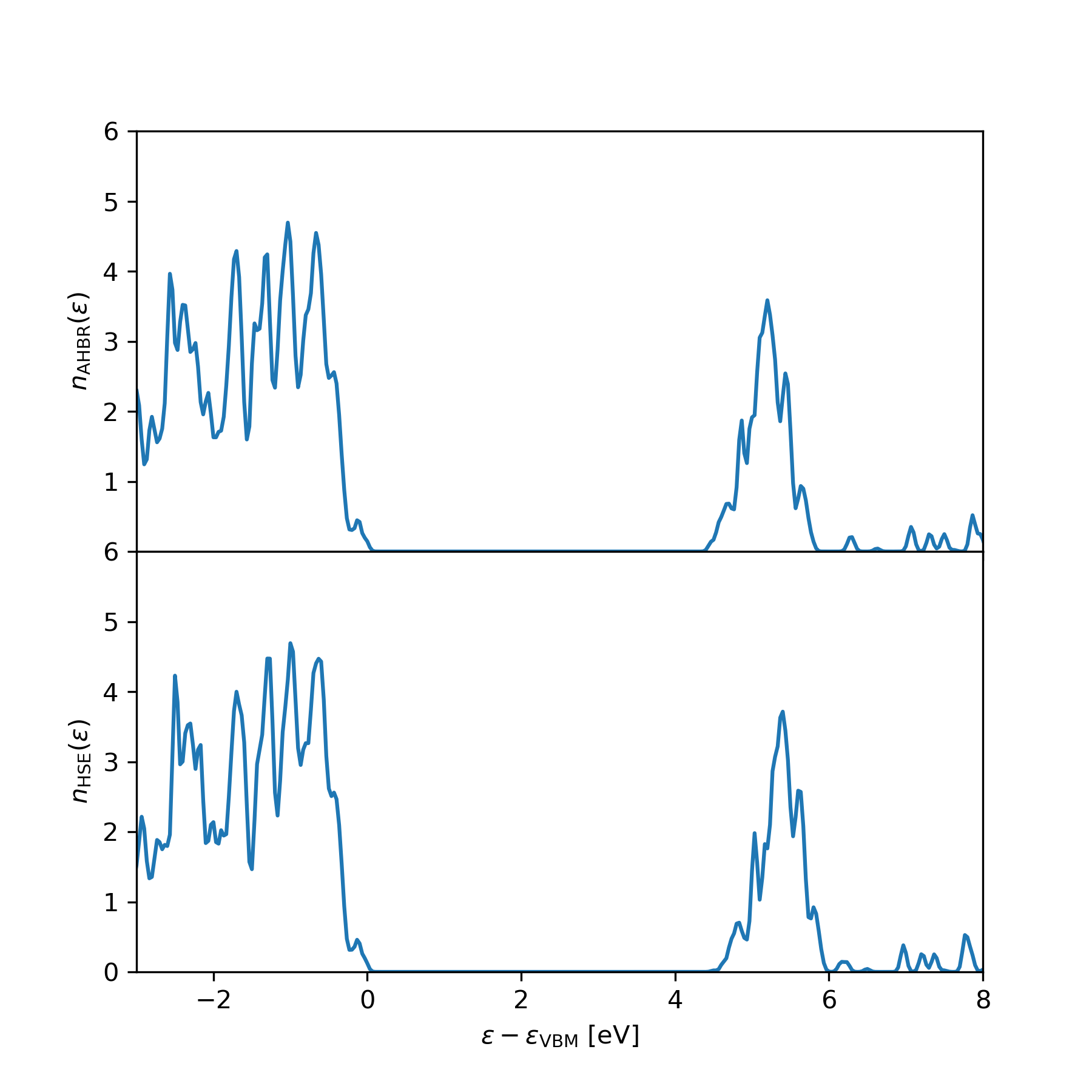}
    \caption{Comparison of AHBR and HSE predictions of DOS variation (top and bottom panels, respectively) for RSAF2 MnO (left) and NiO (right). In each of the comparisons, the shared energy reference is set by the AHBR determination of the VBM. 
    }
    \label{fig:MnONiOrsaf2Bandstructure}
\end{figure*}

\begin{figure}
    \centering
\includegraphics[width=0.95\columnwidth]
{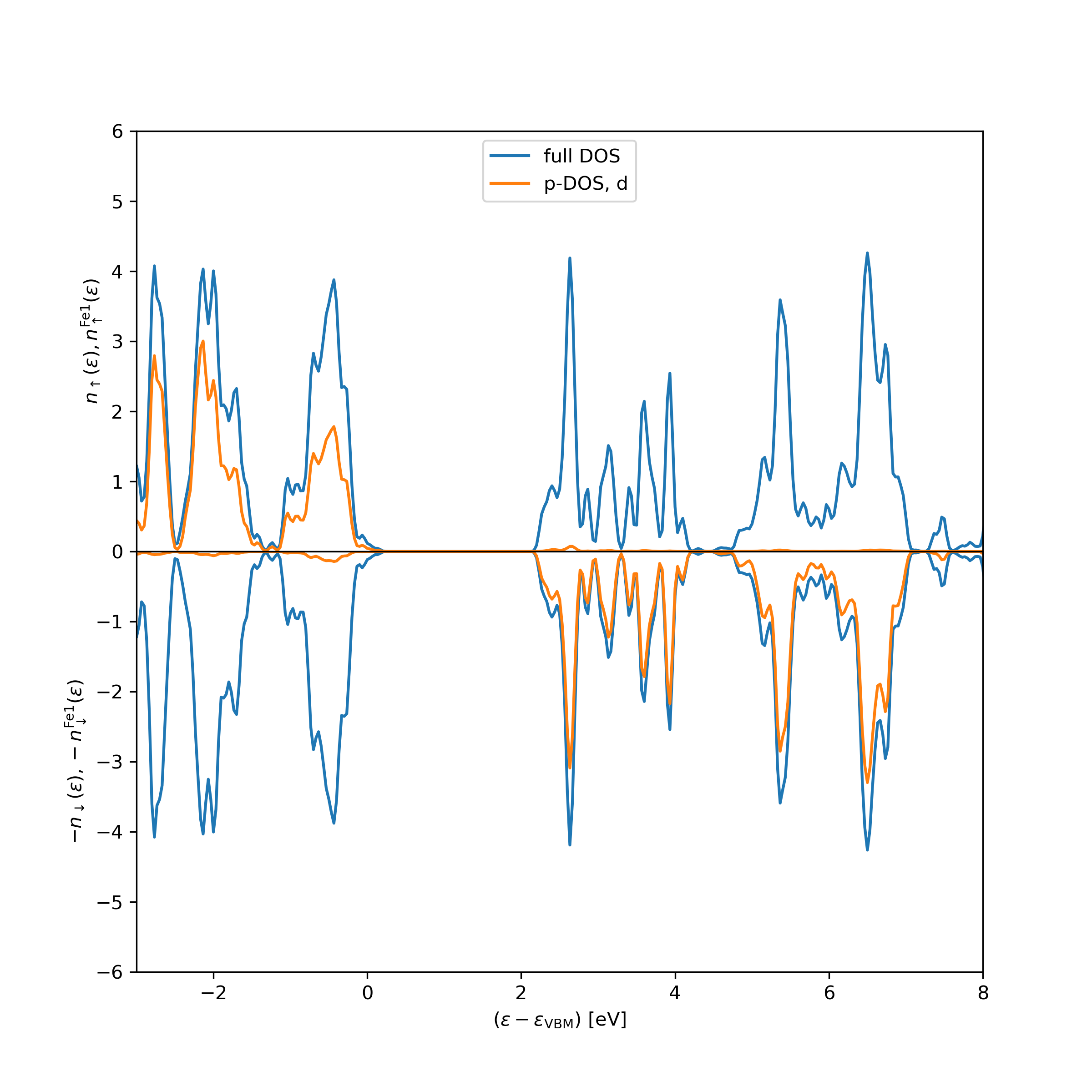}
    \caption{AHBR characterization of the $d$-orbital nature of the
    anti-ferromagnetic-type-II FeO system. The orbitals are filled to the energy of the VBM, i.e., at the onset of the AHBR prediction for the fundamental gap. The panels contrast the total spin-up and spin-down DOS components (blue curves, denoted `full DOS') with their projections onto  the $d$ orbitals of one of the two Fe ions (orange curves, denoted `p-DOS, d'). This subset of $d$-orbitals dominate in setting the occupied (empty) orbitals for spin-up (spin-down) 
    electrons; The corresponding predictions onto the $d$-orbitals
    of the other Fe ion gives exact the opposite variation in such 
    'p-DOS, d' curves.
    }
    \label{fig:FeOrsaf2Bandstructure}
\end{figure}

\subsubsection{Structure and cohesive energies: Variable-cell optimization versus energy variations}

Figure \ref{fig:CROSScheckConverge}  
details our validation of the here-described AHBR-VASP as 
a reliable stress-based predictor of both the lattice 
constants (left panel) and of the associated
cohesive energy (right panel); Appendix B 
summarizes the underlying numerical results. The figure 
panels present a per-bulk-system characterization of relative deviations arising with the present AHBR-QE \cite{AHBRlaunch}
one-parameter (or one-dimensional) manual structure 
optimization (ordinates) and that available by the
new AHBR-VASP studies (co-ordinates). Here
(in this method documentation and illustration work) our interest is not whether a perfectly-converged (and all-electron \cite{Tran19}) 
implementation of AHBR fully aligns with the reference data, i.e.,
$\delta a/a=\delta E/E=0$. Rather, the 
panels focus on showing that we have at least some
instances of near-perfects correlation between 
well-converged AHBR-QE and AHBR-VASP descriptions
of simple bulk problems. That is, when the representations of deviation 
in AHBR-QE and AHBR-VASP characterizations falls on the
dashed diagonal lines, then we can say that AHBR-QE and 
AHBR-VASP studies are making the same error (under full 
convergence), in spite of them using fundamentally different
(manual variation \cite{ziambaras03p064112} versus stress-based \cite{Hard2Soft}) optimization frameworks. 

In practice we complete the argument for here providing
a successful AHBR-VASP implementation as follows. First, we 
see that there are a fair number of bulk systems where the correlation 
mapping falls on the diagonal lines, i.e., having the same 
deviations in VASP as in QE. Noting that AHBR errors on the 
cohesive energy are generally larger than those for 
the lattice constants \cite{AHBRlaunch}, we identify  
such examples as the Au, Al, C, MgO, Si, SiC, and perhaps 
even Rh and Pt systems. There are clear residual
deviations, especially  on the cohesive-energy predictions, 
and we tentatively interpret these as an effect of replacing 
a PP-based AHBR description with the new option for using 
AHBR with PAW setups instead of PPs. However, we can expect that
QE and VASP (both giving approximations to a future 
all-electron AHBR implementation, e.g., based on
Ref.\ \onlinecite{Tran17}) to often or at least sometimes
line up in their materials predictions. Accordingly, 
we find that the good correlation in the Au, Al, C, MgO, Si and SiC
cases suggests that the stress-based optimization for AHBR works, via the AHBR-VASP work.  

Second, the panels of Fig.\ \ref{fig:CROSScheckConverge}
also suggest that there is generally benefit of using the 
AHBR-VASP: We are, in fact, 
gaining an AHBR-PAW-VASP  implementation. The PAW benefits come in terms 
of accuracy in predictions: The 
data points tend to stay closer to the horizontal than to 
the vertical lines, although there are exceptions. For the 
set of bulk test systems with no AHBR-QE/AHBR-VASP correlation, 
it is generally the  AHBR-PAW-AHBR predictions that lower the 
deviation from the benchmark info \cite{Tran19}. This observation 
suggests that use of PAW setups (in VASP) rather than PPs (in AHBR-QE) can 
generally help the accuracy. The systematics of this 
improvements-by-PAW trend also suggests (as anticipated) 
that we can ascribe some of the residual deviations in AHBR characterizations
(at least for Si, SiC, LiF, NaCl, Al, and Pt) to a question of PP usage.

\subsection{Using AHBR-VASP for discovery}

We finally turn to illustrations of the 
AHBR-VASP capability for serving for materials discovery in practice. Here, we include two types of explorations, namely 1) cases where independent AHBR predictions can, in part, eventually be tested by comparison against existing experimental observation combined with other trusted theory, and 2) cases where we are making genuine predictions so that our work can later, independently be validated or used as input to seek further XC improvements.

\subsubsection{Metal-monoxide structure and bandgaps}

We pick our metal-oxide unit-cell modeling so that it is possible for the unit-cell representation to support the antiferromagnetic order of type II; This is done by doubling the basic rock-salt unit cell, as shown in
Fig.\ \ref{fig:SchematicsVCrelax}.
In making this modeling choice,  we take as prior input the experimental observation that this anti-ferromagnetic order is a characteristic of the ground-state of  MnO, NiO, and FeO. Beyond that input, however, we 
proceed here as if facing a proper discovery challenge. That is, we apply AHBR-VASP with the option of stress-based structure optimization and simply
await the outcome of studies for which
we document convergence, for example, with the choice of energy cut off, Appendix B.

Table \ref{tab:MOsummary} summarizes the AHBR-VASP predictions of structure, cohesion, and magnetization order in MnO, NiO and FeO. We find
that there are small distortions, $a' \neq a$, in what remains an underlying almost-rocksalt
form. The AHBR predictions of the base lattice constants $a$ are found in
good agreement with experimental observations. The small distortions
occur in the direction that carries the anti-ferromagnetic ordering.

\begin{figure}
    \centering
\includegraphics[width=0.27\columnwidth]
{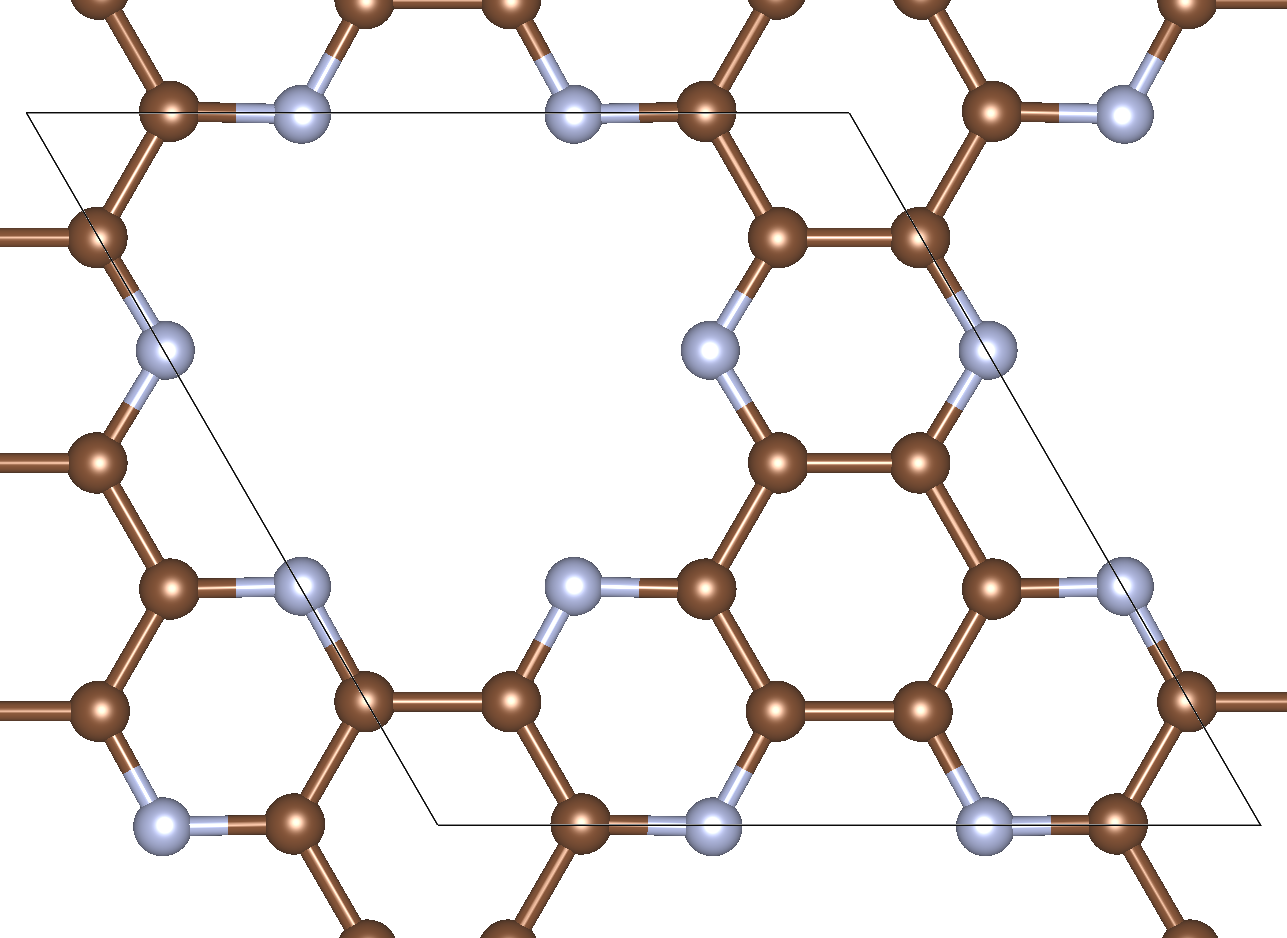}\,
\includegraphics[width=0.33\columnwidth]
{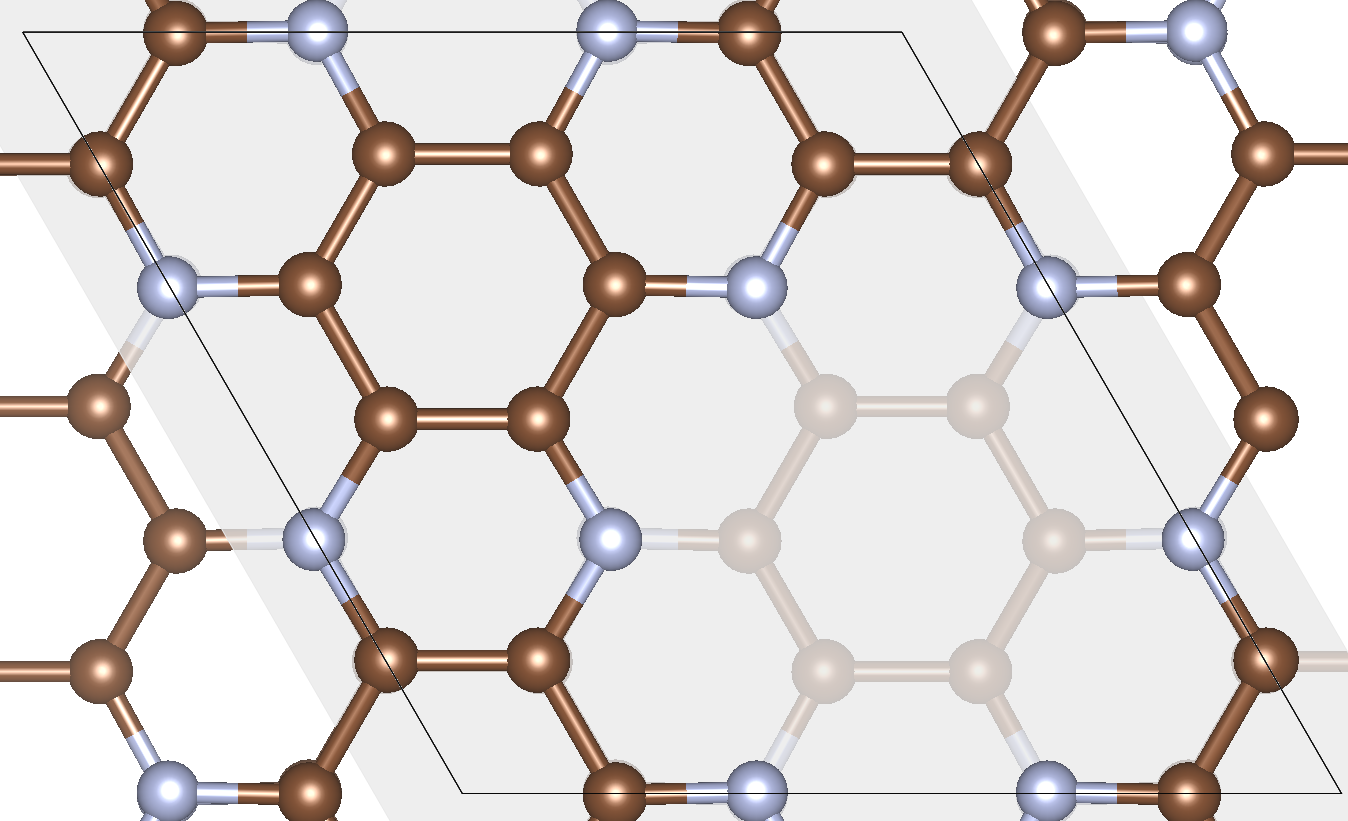} \,
\includegraphics[width=0.33\columnwidth]
{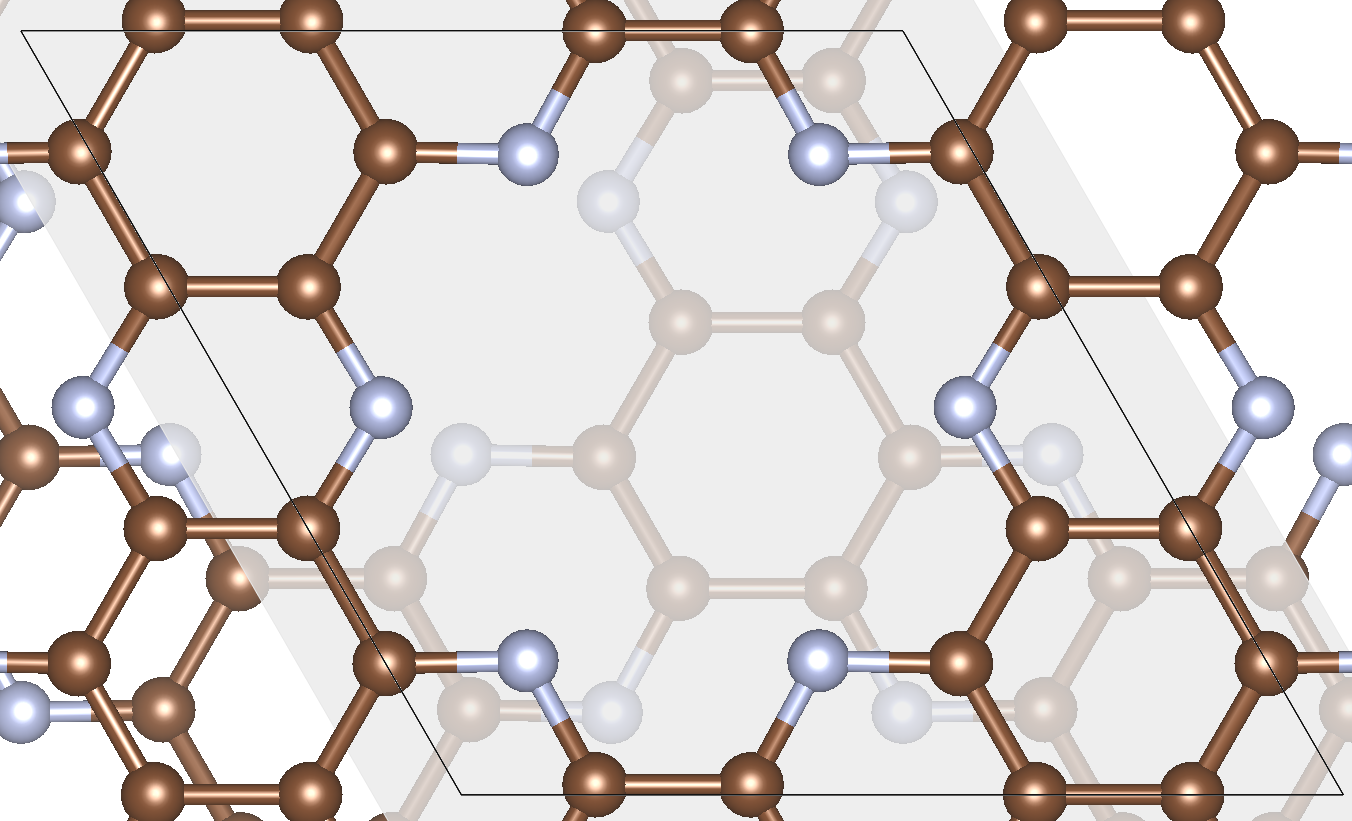}
    \caption{AHBR predictions of
    individual-sheet structure (left) and of most-relevant metastable layer stacking forms (middle and right) of COF-C2N; Blue (brown) balls identify nitrogen (carbon) atoms.  The left panel simultaneously shows the structure of the AA stacking form as viewed along the sheet normal, when atoms in the lower sheets are hidden by those of the top sheet. The following panels contrast the nature of traditionally-investigated AB stacking form (middle) and of the here-identified lower-energy `parallel-displaced (PD) AB stacking motif (right). Here, use of a grey, partly transparent, layer serves to identify which of the visible atoms belong in the lower sheet; Some six-C rings remain `occluded,' that is, situated immediately below a ring in the top sheet in this AB stacking form. The partial ring alignment disappears as the system relaxes to the lowest-energy motif PDAB, see text.
    }
    \label{fig:C2NCOFstruct}
\end{figure}

\begin{figure}
    \centering
\includegraphics[width=0.85\columnwidth]
{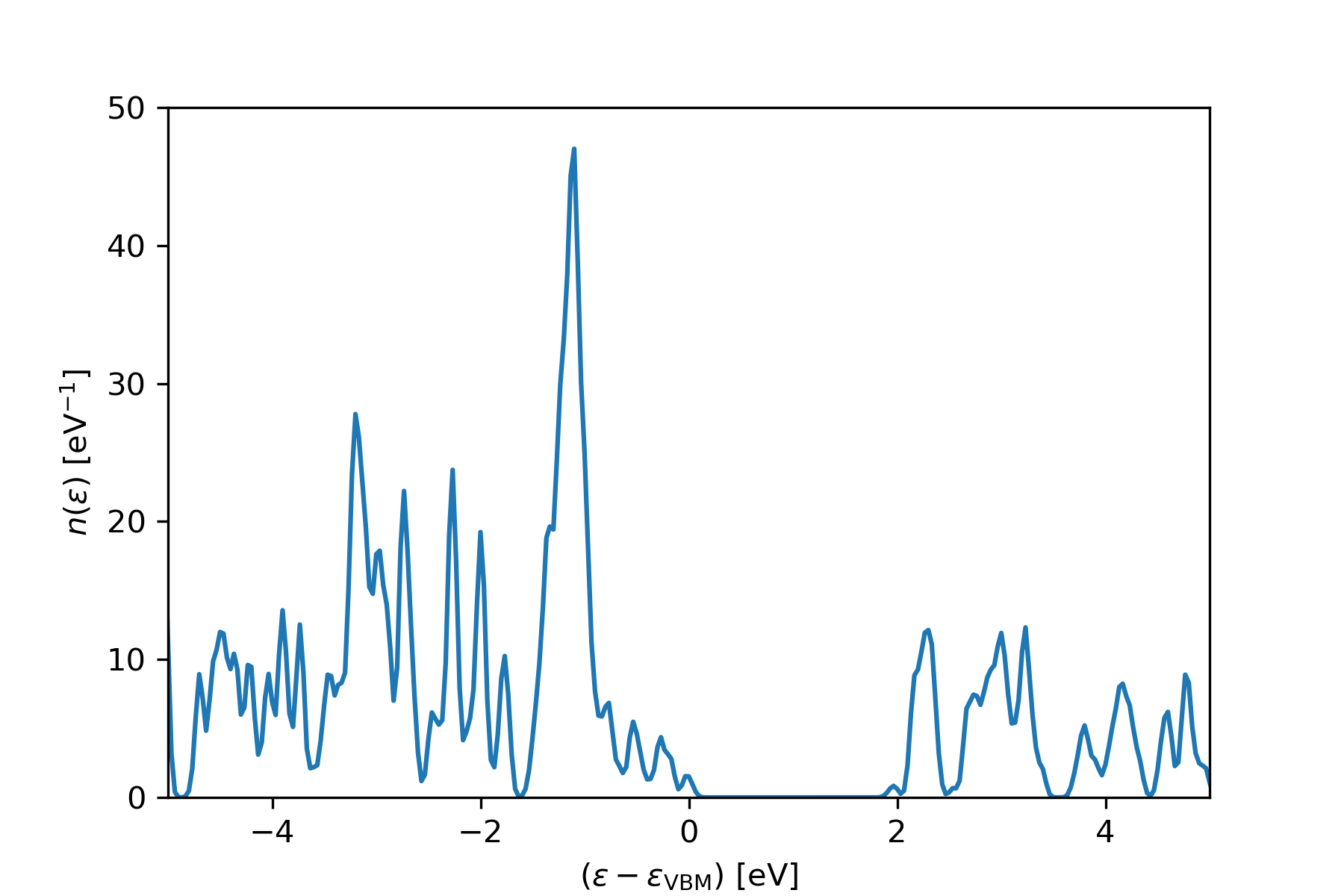}\\
\includegraphics[width=0.85\columnwidth]
{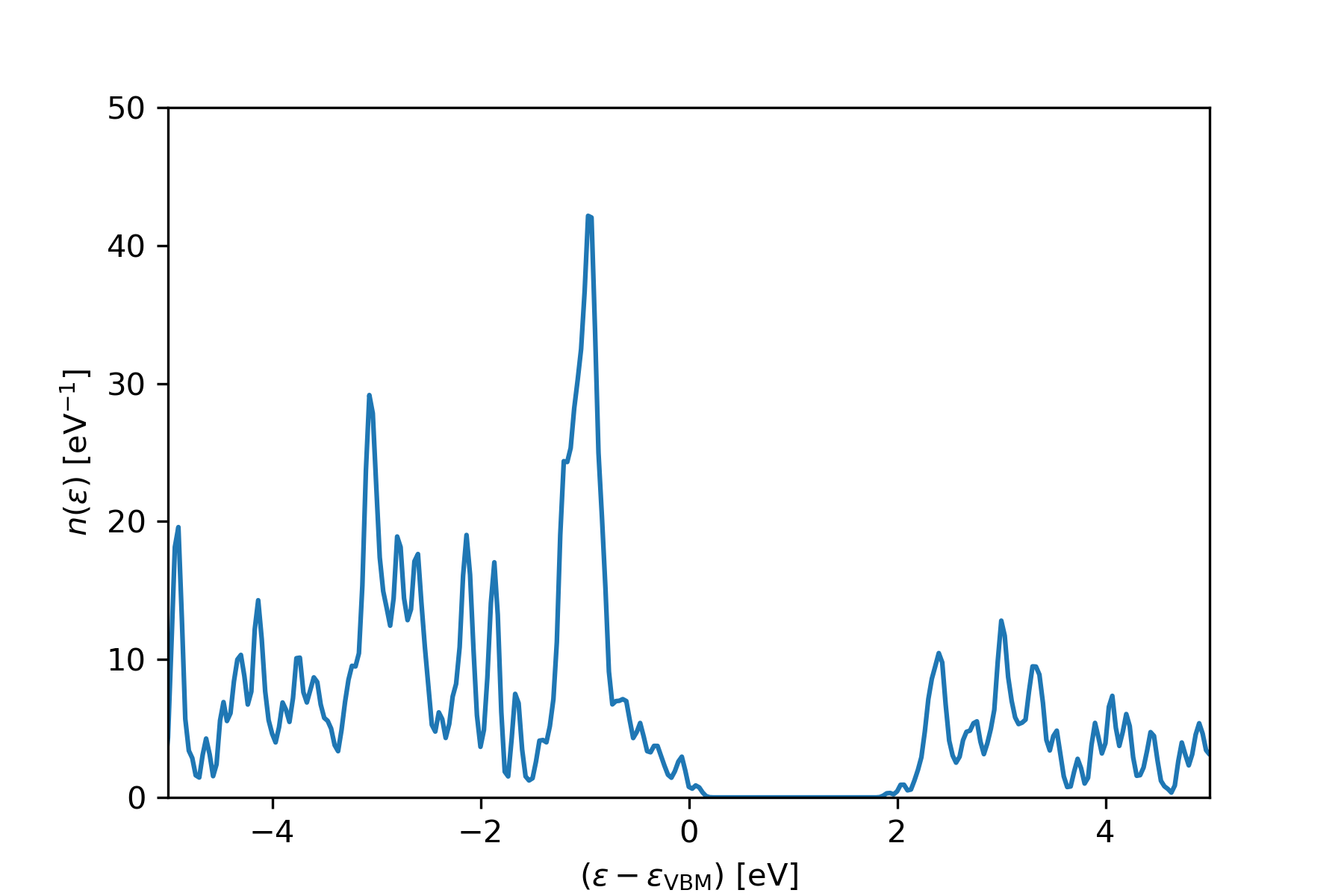}
    \caption{Contrast of DOS characterization, at AHBR-predicted optimal structures for the AB stacking (top) and 
    the PDAB stacking (bottom) of C2N COF.
    }
    \label{fig:C2N-COF-DOS}
\end{figure}

The choice of the metal monoxides
as a first complex-matter testing case is made because these are, after all, systems where other trusted 
theory can be used to assert the structure. Specifically, we may use HSE06 in VASP (at correspondingly defined high accuracy) to check whether the slight symmetry breaking 
is an AHBR-VASP 
implementation artifact or, instead, plausible.

Table \ref{tab:MOsummary} includes a comparison of AHBR-VASP and HSE06-VASP characterizations for MnO and NiO. We find that there is, in fact, 
a close alignment between VASP-based HSE06 predictions and our new AHBR-VASP predictions of structure. A key observation is here that the metal monoxides systems have strong partly ionic bonds and both HSE and AHBR are therefore expected to accurately reflect the extent of charge-transfer effects. They should (and they do) agree on structure predictions because
vdW interactions are not expected to provide significant changes in such dense-matter cases.

The pair of top panels of Fig.\ \ref{fig:MnONiOrsaf2Bandstructure}
summarize our AHBR predictions of the MnO and NiO electronic structure as reflected in predictions for the spin-up density of  state (DOS) variation; These variations are exactly the same as the spin-down variations (not shown), reflecting perfect anti-ferromagnetic ordering. Table \ref{tab:MOsummary}
summarizes the results that we extract for the fundamental gap, when ignoring DOS contributions smaller than 0.1 eV$^{-1}$. The predictions for $\Delta_g$ align with experimental characterizations 
for MnO, NiO and FeO. There is also alignment with corresponding HSE06
\cite{HSE06} determinations for $\Delta_g$ (obtained at the structure 
determined instead by HSE), Table
\ref{tab:MOsummary}.

Moreover, our DOS focus and our  HSE06 characterizations for MnO and NiO allow us to further document that the AHBR itself (in AHBR-VASP coding) yields plausible electronic-structure predictions. The HSE06 DOS results are reported (again for the spin up component) in the set of
bottom panels in Fig.\ \ref{fig:MnONiOrsaf2Bandstructure}.
There are in both metal-oxide cases
a good qualitative agreement with the AHBR DOS prediction. This is, again, expected because vdW forces are not expected to have a dramatic impact on the MnO and NiO behavior.

Figure \ref{fig:FeOrsaf2Bandstructure}
reports our work to further detail the FeO electron structure,  as revealed in total and projected DOS components. The top and bottom  panels now contrast the spin-up and spin-down DOS components
both in terms of the overall DOS (blue curves) and in terms of the projections onto the $d$ orbitals of one of the two Fe atoms (orange curves). The AHBR-VASP characterizations retain a spin-up and spin-down symmetry for the total DOS. However, when we project onto
the subset of $d$-orbitals that reside on either one of the Fe ions, the nature of the anti-ferromagnetic-type-II order is revealed: We have the d-band
originating at the metal atoms mostly occupied and mostly unoccupied on every other Fe (or Mn or Ni) atoms in the unit cell,
Fig.\ \ref{fig:SchematicsVCrelax}.
Meanwhile, the metal $d$ orbitals do not produce any net or overall magnetization due to the spatially alternating nature of $d$-band contributions to the occupied part of the system DOS. 

\begin{table}
\caption{\label{tab:C2NCOFsummary}
Results of VASP-based vdW-DF-cx (denoted `CX') and AHBR-VASP (`AHBR') variable-cell optimization of the
hexagonal lattice constants: $a$ and $c$ lattice constants, unit-cell volume $\Omega$ and binding energy $E_{\rm bind}^{\rm cell}$, for C2N-COF stacking motifs, i.e., plausible, metastable forms for stacking in this layered system. 
Italized entries identify the lowest-energy motif, i.e., our prediction for a possible ground-state stacking form. Also shown are corresponding predictions of the `total helium volume' (THV) and 'pore limiting diameter' (PLD). The former (latter) characterizes the space, per unit cell, that is available for incorporating a small atom/ion,  an alkali dopant, without driving significant structural changes in the set of metastable C2N-COF stacking forms (the maximum-size condition for
inhibiting even neutral-object diffusion once 
synthesis is completed). THV and PLD predictions are given in cm$^3$/g and {\AA}, respectively.}
\begin{tabular}{ll|ccc|ccc}
\hline
\hline 
Motif & XC & $a$ [{\AA}] & $c$ [{\AA}] & $\Omega$ [{\AA}$^{3}$] & $E_{\rm bind}^{\rm cell}$ & THV & PLD \\
\hline
AA & CX & 8.308  & 7.131  & 426  &  -1.283 & 0.055 & 2.16 \\
& AHBR & 8.251 & 7.062  & 416 & -1.028 & 0.053 & 2.14\\ 
 AB & CX & 8.303  & 6.404  & 382 &  -2.280 & 0.007 & 0.80   \\ 
& AHBR & 8.246 & 6.401 & 377 & -2.103 & 0.006 & 0.78 \\
 \textit{PDAB} & CX & 8.301  & 6.248  & 373  &  \textit{-2.378} & 0.005 & 0.77 \\ 
& AHBR & 8.245 & 6.283  & 370 & \textit{-2.188} & 0.004 & 0.77 \\
\hline
\hline
\end{tabular}
\end{table}

\subsubsection{Stacking and electronic structure of an incompletely characterized COF system: C2N}

Figure \ref{fig:C2NCOFstruct} shows atomic structure of 
the C2N-COF systems in various metastable stacking structures. The left panel shows the atomically-flat individual-sheet structure which is formed by (all-carbon) aromatic rings linked by a pair of nitrogen atoms. The individual sheet resembles graphene although with rings instead of 
isolated carbon-atom vertices and wider, two-nitrogen-atom linkers
instead of a C-C bond. The panel also shows a schematics of the layered three-dimensional structure in the AA stacking form that is metastable and that again has a
hexagonal unit cell (thin solid lines); see
Table \ref{tab:C2NCOFsummary} for summary of 
AHBR-VASP predictions for lattice parameters. 
The panel shows the AA stacking from a direction
that is on the normal of the layer plane. As such the panel 
also reveals the nature of the (here-connected) internal 
atom holes: Benzene-size voids bordered by nitrogen with 
some potential reactivity and hence options for engineering 
the DOS of this semiconducting COF.

The middle panel shows the unit-cell form in the so-called (layered) AB stacking form that can be viewed as a form of super-sized graphite, i.e., a hexagonal cell with a staggering of the sheets. The lateral position of the atom voids of individual sheets no longer align. In this metastable AB form we now only
stack one-half of all aromatic-ring vertices, and one-half the nitrogen-atom linkers, on top of each other (unlike in the AA case). This is important (here, as in graphite)
since the carbon rings have, as aromatic structures, large quadrupole moments
\cite{NitrogenBasesAHBR-mRSH26}. The interplay of electrostatics and vdW attraction,
means, for example, that the optimal structure of the benzene dimer involves parallel displaced  (PD) stacking. Here we document a similar impact for C2N, finding 
that the AB stacking motif is energetically favored over the AA stacking by more 
than 1 eV per unit cell.

The right panel of Fig.\ \ref{fig:C2NCOFstruct} shows
the structure of what we predict as the optimal stacking, namely 
what we call a PD AB form (abbreviated PDAB). This holds
whether we use CX or AHBR for our stress-based structure optimization.
The structures have atomically flat sheets, remain hexagonal, given by the lattice-constant predictions summarized in Table \ref{tab:C2NCOFsummary}. However, there is here a small additional lateral shift among the layers, corresponding to an additional energy gain (about 33 meV per unit cell)
compared with the AB stacking form. We interpret this 
energy optimization again in terms of the competition between vdW interlayer attraction, steric hindrance,  and electrostatic effects (as in the benzene dimer). 
While the AB form  got rid of half of the suboptimal
direct-stacking cases, the PDAB completes the
structure optimization by further 
implementing consequences of the competition between inter-sheet electrostatics coupling and vdW attraction.  For example, the unit-cell height $c$ is lowered (thereby strengthening the vdW binding contribution \cite{JiScHy18a}) by this quadrupole- (or PD-)type driving mechanism in the PDAB form, Table \ref{tab:C2NCOFsummary}.

Figure \ref{fig:C2N-COF-DOS} contrasts the AHBR-VASP  predictions of the C2N COF DOS as evaluated at the
AHBR predicted structures for the AB and PDAB 
forms, top and bottom panels, respectively. The
figure shows that the AHBR characterizations
find a fundamental or electronic gap,  
$\Delta_g\approx 2$ eV, that is essentially 
unchanged by the PD shift. This band gap value 
is in good agreement with experimental observation 
for this semiconducting COF system \cite{C2Nstart2015,LeeGapNanoPore2025}.

Table \ref{tab:C2NCOFsummary} also makes it plausible that alkali-metal doping of C2N can succeed if pursued at the synthesis stage. The table reports our analysis of the nature of C2N pores, completed by use of the \textsc{Mercury} and \textsc{PoreBlazer} codes \cite{MercuryI,MercuryII}, across the key metastable forms or motifs for C2N-COF stacking. The analysis is given in terms of predictions for the so-called total helium volume (THV) and for the effective so-called pore-limited diameter (PLD). The THV measure reflects an assumption that that the (inert) He atom has a vdW radius of about  1.4 {\AA} and reveals the extent that the Pauli exclusion leaves
room for such objects inside the C2N COF, in the set stacking motifs.
Meanwhile, the essentially vanishing value of the PLD measure implies that 
He atom will be unable to escape: The would-be C2N-He system is technically here
predicted to be a metastable He-based layered compound, assuming again that 
such doped C2N can be synthesized in the first place.

There are two conclusions of potential relevance that can be extracted by our predictions for the semiconducting
C2N system. As a background, we observe 
that the Na ion has a vdW radius 
$r_{\rm Na+}$ that is expected to be about 10\% smaller \cite{ScharffAtomSize} 
than that of the Ne atom, $r_{\rm Ne}\approx 1.54$ {\AA} \cite{Bondi1964}, and therefore comparable in size to the He atom,  $r_{\rm Na+} \approx r_{\rm He} \approx 1.40$ {\AA} \cite{Bondi1964}. The first conclusion is that our AHBR-VASP predictions for the optimal C2N structure and associated THV, Table \ref{tab:C2NCOFsummary}, suggest that one can insert sodium atoms as dopants, essentially without creating strain: The voids in the C2N-COF sheets are already in place. The second conclusion is that such Na
doping is expected to be stable: The predicted
PLD value of the optimal stacking PDAB, Table \ref{tab:C2NCOFsummary}, is much smaller than $r_{\rm He}$ and $r_{\rm Na+}$. The conclusion of stable Na-atom inclusion (assuming these were inserted at the synthesis stage) holds even if we were to assume that the Na atoms or ions are as inert as helium (which obviously they are not).

\begin{table}
    \caption{ 
Binding energies (in kcal/mol) of the S22 benchmark dataset, calculated using the AHBR functional with 0.25 Fock-exchange mixing
and $\gamma=0.2$ {\AA}$^{-1}$.
Comparison of VASP, QE (quantum espresso) and GMTKN55 values (obtained in SAPT2+3(CCD)/aug-cc-pVTZ) \cite{gmtkn55}. \label{tab:tabelS22}}
    \centering
    \begin{tabular}{c c c c}
\hline
S22 compound & AHBR-VASP & AHBR-QE &  Ref. \\
\hline 
 \multicolumn{4}{c}{H-bonded systems} \\
\hline
1 & 2.909 & 2.963 & 3.133 \\
2 & 4.949 & 5.026 & 4.989 \\
3 & 19.416 & 19.622 & 18.753 \\
4 & 15.980 & 16.093 & 16.062 \\
5 & 20.404 & 20.550 & 20.641 \\
6 & 16.836 & 16.991 & 16.934 \\
7 & 16.071 & 16.224 & 16.66 \\
\hline 
 \multicolumn{4}{c}{Dispersion systems} \\
\hline
8 & 0.512 & 0.517 & 0.527 \\
9 & 1.224 & 1.256 & 1.472 \\
10 & 1.274 & 1.290 & 1.448 \\
11 & 2.283 & 2.272 & 2.654 \\
12 & 3.693 & 3.721 & 4.255 \\
13 & 9.573 & 9.626 & 9.805 \\
14 & 3.811 & 3.808 & 4.524 \\
15 & 11.078 & 11.148 & 11.73 \\
\hline
 \multicolumn{4}{c}{Mixed systems} \\
\hline
16 & 1.562 & 1.607 & 1.496 \\
17 & 3.063 & 3.114 & 3.275 \\
18 & 2.106 & 2.140 & 2.312 \\
19 & 4.361 & 4.461 & 4.541 \\
20 & 2.410 & 2.450 & 2.717 \\
21 & 4.967 & 5.022 & 5.627 \\
22 & 6.525 & 6.594 & 7.097 \\
\hline
MD & -0.257& -0.189 & \\
\hline
    \end{tabular}
\end{table}

\section{Summary}

We have implemented AHBR in VASP and have thereby also gained access to
robust PAW-based materials characterizations by this recent RSH vdW-DF.
Importantly for the discovery-by-AHBR focus that we also pursue, we have 
thereby also enabled the use of AHBR for robust stress-based structure 
optimization for (soft and other) complex materials.

We have tested our QE-to-VASP porting by comparison of the
performances on simple bulk systems and for slightly distorted systems, namely the
anti-ferromagnetic metal monoxides MnO, NiO, and FeO. The general alignment 
between well-converged AHBR-VASP and AHBR-QE structure predictions and the finding of acceptable deviations by such AHBR description (relative to the so-called Solid-44 benchmark set) is promising. It is seen as confirmation that we have correctly implemented both 
this RSH vdW-DF, gotten it to work for stress-based optimization in VASP, and
an indication of overall method usefulness. 
The implementation and stress-based optimization is also tested
on predicting the atomic structure and DOS signatures in the slightly distorted rocksalt metal monoxide systems. Here our AHBR predictions are validated by 
comparison with VASP-HSE06 studies.

Finally, the use of AHBR-VASP as a tool for discovery is demonstrated by using it to 
predict the three-dimensional stacking of the individual sheets in the C2N COF. 
This stacking is set by a complex interplay between the steric hindrance, vdW attraction, and the electrostatics set by the aromatic-type building blocks 
forming the individual C2B sheets. Accuracy in predicting possible charge transfer
is therefore essential but this is also a challenge where the AHBR has previously
been successful (as
shown in by AHBR-QE studies. 

We note that the C2N COF does not have a complete experimental characterization when it comes to understanding the three-dimensional stacking, there is just a discussion of simple motifs for this stacking.  Our theory exploration of the C2N COF may later be confirmed or proven incorrect by  experiments but it is, in itself, relevant to show that the new tool can make predictions for subsequent assessments.

\section*{Acknowledgement}
Work supported  by Swedish Research Council (VR) through Grant No.\ 2022-03277 (PH), 
by the Chalmers Area of Advance (AoA) Nano and Chalmers AoA Production (PH and L{\"O}), and by Chalmers e-Commons via assistance with code porting and implementation (YS and PH).
The authors furthermore acknowledge support in the form of computational and storage resources at 
Chalmers Centre for Computational Science and Engineering (C3SE),
and from the 
National Academic Infrastructure for Supercomputing in Sweden (NAISS),  within computing and storage projects
NAISS2023/3-22, 
NAISS2023/6-306, 
NAISS2024/3-16,   NAISS2024/6-432,
and NAISS2025-3-25.

\appendix

\section{S2 binding-energy differences}

Table \ref{tab:tabelS22} summarizes and contrasts AHBR-VASP and AHBR-QE results for the set of non-covalent
interaction problems of the S22 benchmark set \cite{gmtkn55}. All values are here in kcal/mol. Reference geometries and
reference binding energies
(listed for comparison and allowing an assessment of AHBR accuracy), are taken from Ref.\ \onlinecite{gmtkn55}. The energies reflect calculations completed with symmetry-adapted coupled-cluster calculations, again as summarized in Ref.\ \cite{gmtkn55}.

We note that, in contrast to the AHBR (or AHBR-QE) assessment given in Ref.\ \cite{AHBRlaunch}, we
have here omitted use of an explicit electrostatic decoupling between cells of repeated images of molecules and molecular systems \cite{Makov}. This is done
since the focus is here on the
AHBR-VASP and AHBR-QE comparison and 
it was not clear that we could get exactly the same type of explicit
decoupling in the two codes. We have however, in both testa proceeded to use a large unit-cell representation \cite{AHBRlaunch}.

\section{Convergence  of AHBR-PAW-VASP and AHBR-PP-QE for bulk systems}

\begin{table}[h]
\caption{\label{tab:SimpleBulkMethodCompareL} Lattice constants $a$ of simple bulk systems at various 
$k$-point and $q=\Delta k$-point samplings (and at ENCUT=1600 eV) for a selection of simple metal and nonconductor cases. The materials represent
the set on which we subsequently validate the AHBR-VASP
variable-cell optimization by cross-checking accuracy
with AHBR-QE results. We list reference experimental 
values as back-corrected for vibrational effects, taken from 
Ref.\ \cite{Tran19}. Italicized entries reflect AHBR results obtained at the accuracy level that we considered converged for both AHBR-VASP and AHBR-QE characterizations, namely
$(k,q)=(12,6)$ and $(k,q)=(8,8)$ for metals and nonconductors, respectively. 
}
\begin{tabular}{l|cc|c|cc}
\hline
\hline
System
& 
VASP-8-8
& 
VASP-12-6
& 
Exper.
&
QE-8-8
&
QE-12-6
\\
\hline
Al  & 4.029 & \textit{4.029} &  4.033 & 4.034 & \textit{4.036} \\
Cu  & 3.637 & \textit{3.626} &  3.599 & 3.615 & \textit{3.607} \\
Ag & 4.141 & \textit{4.130} & 4.070 & 4.116 & \textit{4.108} \\
Au & 4.131 & \textit{4.125} &  4.067 & 4.125 & \textit{4.118} \\
Pt & 3.923 & \textit{3.928} &  3.917 & 3.924 & \textit{3.931} \\
Rh & 3.781 & \textit{3.776} &  3.786 & 3.769 & \textit{3.765} \\
\hline
C & \textit{3.545} & 3.545 &  3.553 & \textit{3.544} & 3.544 \\
Si & \textit{5.429} & 5.429 &  5.411 & \textit{5.441} & 5.441 \\
SiC & \textit{4.344} & 4.344 &  4.346 & \textit{4.351} & 4.351 \\
GaAs & \textit{5.675} & 5.675 & 5.638 & \textit{5.639} & 5.640 \\
MgO (RS) & \textit{4.192} & 4.191 &  4.189 & \textit{4.187} & 4.187 \\
LiF  & \textit{3.983} & 3.983 &  3.972 & \textit{3.996} & 3.997 \\
NaCl & \textit{5.598} & 5.599 &  5.569 & \textit{5.597} & 5.597 \\
\hline
\end{tabular}
\end{table}

Tables \ref{tab:SimpleBulkMethodCompareL} 
and \ref{tab:SimpleBulkMethodCompareE} summarize the AHBR-VASP and AHBR-QE system characterizations for lattice constants $a$ and cohesive energy $E_{\rm coh}$, respectively. Predicted values are reported (in {\AA} and eV) at the two major options for convergence in $k,q$ values; Italicized entries reflect results that we consider converged for metals (top table parts) and for nonconducting systems (bottom table parts). We note that the same conditions are found sufficient for 
convergence in both implementations so we can now directly compare performance of each of these with regards to the bulk-system reference 
data \cite{Tran19}.

\begin{table}
\caption{\label{tab:SimpleBulkMethodCompareE} AHBR predictions of the cohesive energies $E_{\rm coh}$ (in eV) at indicated 
$k$-point and $q=\Delta k$-point samplings. As highlighted again we find a choice of 12-6 and 8-8 $k-q$ combinations sufficient for metals and semiconductors, respectively.
Experimental values are back-corrected for vibrational effects, taken from Ref.\ \cite{Tran19} and included for reference. 
}
\begin{tabular}{l|cc|c|cc}
\hline
\hline
System
& 
VASP-8-8
& 
VASP-12-6
& 
Exper.
&
QE-8-8
&
QE-12-6
\\
\hline
Al (fcc) & 3.296 & \textit{3.315} & 3.431 & 3.200 & \textit{3.208} \\
Cu (fcc) & 3.032 & \textit{3.067} & 3.513 & 3.063 & \textit{3.086} \\
Ag (fcc) & 2.507 & \textit{2.536} & 2.964 & 2.549 & \textit{2.566} \\
Au (fcc) & 3.164 & \textit{3.199} & 3.835 & 3.159 & \textit{3.182} \\
Pt (fcc) & 5.276 & \textit{5.346} & 5.866 & 4.949 & \textit{4.990} \\
Rh (fcc) & 4.897 & \textit{4.938} & 5.783 & 4.645 & \textit{4.695} \\
\hline
C  & \textit{7.657} & 7.658 & 7.452 & \textit{7.380} & 7.361 \\
Si & \textit{4.694} & 4.696 & 4.685 & \textit{4.556} & 4.555 \\
SiC & \textit{6.498} & 6.609 & 6.478 & \textit{6.281} & 6.278 \\
GaAs & \textit{3.260} & 3.261 & 3.337 & \textit{3.188} & 3.188 \\
MgO (RS) & \textit{5.045} & 5.043 & 5.203 & \textit{4.947} & 4.945 \\
LiF & \textit{4.404} & 4.389 & 4.457 & \textit{4.393} & 4.392 \\
NaCl  & \textit{3.247} & 3.246 & 3.337 & \textit{3.193} & 3.193 \\
\hline
\end{tabular}
\end{table}

Specifically, the metal-monoxides have just an almost-rocksalt structure, meaning that there is still a main rocksalt lattice 
constant $a$ but also a small distortion 
$a'/a \neq 1$. The latter arises in  the direction that carries the 
type-II anti-ferromagnetic order,
see also Fig.\ \ref{fig:SchematicsVCrelax}.
For these lower symmetry problems, Table \ref{tab:EcutConvergeVASP} investigates which ENCUT value suffices. This is done by supplementing the convergence 
testing by tracking the
variation in predicted $a'/a$ values. We furthermore add a simple qualitative test on the AHBR-VASP descriptions: Is the predicted magnetic order exclusively anti-ferromagnetic?

\begin{table}
\caption{\label{tab:EcutConvergeVASP} Convergence of VASP variable-cell structure determinations 
with regards to the energy cut off. This convergence is documented at the choice of $k$-point and $q$-point samplings that we 
generally deem more than sufficient for these examples of metal and semiconducting systems. Lattice constants $a$ in {\AA} and cohesive-energies $E_{\rm coh}$ in eV, where available. Exclamations marks warn that we need a high ENCUT value to get magnetization of the Ni atom and therefore a relevant prediction for the NiO $E_{\rm coh}$ value. 
}
\begin{tabular}{l |cc c| cccc}
\hline
\hline
System & \multicolumn{2}{c}{Sampling} & ENCUT
& $a$ & $E_{\rm coh}$ & AF2 & $a'/a$ \\
\hline
Cu & $k=12$ & $q=12$ & 1200 & 3.628
& 3.063 & - & -
\\
Cu & $k=12$ & $q=12$ & 1400 & 3.628
& 3.062 & - & - 
\\
Cu & $k=12$ & $q=12$ & 1600 & 3.628
& 3.062 & - & -
\\
\hline
C & $k=12$ & $q=12$ & 1200 & 3.545
& 7.657 & - & -
\\
C & $k=12$ & $q=12$ & 1400 & 3.545
& 7.657 & - & -
\\
C & $k=12$ & $q=12$ & 1600 & 3.545
& 7.658 & -& -
\\
C & $k=8$ & $q=8$ & 1600 & 3.545
& 7.657 & - & -
\\
SiC & $k=8$ & $q=8$ & 1200 & 4.344  
& 6.497 & - & -
\\
SiC & $k=8$ & $q=8$ & 1400 & 4.344 
& 6.498 & - & -
\\
SiC & $k=8$ & $q=8$ & 1600 & 
4.344 & 6.498 & - & -
\\
\hline
NiO & $k=8$ & $q=8$ & 1200 & 4.166
& ! & yes & 0.998
\\
NiO & $k=8$ & $q=8$ & 1400 & 4.166
& ! & yes & 0.998
\\
NiO & $k=8$ & $q=8$ & 1600 & 4.167
& 4.542 & yes & 0.998
\\
FeO & $k=8$ & $q=8$ & 1200 & 4.344
& 4.673 & yes & 0.979
\\ 
FeO & $k=8$ & $q=8$ & 1400 &  4.341 & 4.672 & yes & 0.980
\\
FeO & $k=8$ & $q=8$ & 1600 & 4.342
& 4.673 & yes & 0.979
\\
\hline
\end{tabular}
\end{table}

\vfill
\eject

%


\end{document}